\documentclass[journal,compsoc]{IEEEtran}

\usepackage{booktabs} 
\usepackage{graphicx}
\usepackage{epstopdf}
\usepackage{subfigure}
\usepackage{diagbox}
\usepackage{multirow}
\usepackage{amsfonts}
\usepackage{subfigure}
\usepackage{amsmath}
\usepackage{enumerate}
\usepackage{amssymb, bm, amsthm}
\usepackage[ruled,vlined,linesnumbered]{algorithm2e}

\usepackage[nocompress]{cite}
\usepackage[numbers,sort&compress]{natbib}

\ifCLASSOPTIONcompsoc
  \usepackage[nocompress]{cite}
\else
  \usepackage{cite}
\fi
\ifCLASSINFOpdf

\else

\fi

\usepackage{array}

\begin{document}

\title{Deep Pairwise Hashing for Cold-start Recommendation}

\author{Yan Zhang, Ivor W. Tsang, Hongzhi Yin, Guowu Yang, Defu Lian, and Jingjing Li
\IEEEcompsocitemizethanks{
\IEEEcompsocthanksitem Y. Zhang, is with the School of Computer Science and Engineering, University of Electronic Science and Technology of China, Chengdu 611731, China, and the Centre for Artificial Intelligence, University of Technology Sydney, Sydney 2007, Australia.  \protect\\ E-mail: yixianqianzy@gmail.com. 
\IEEEcompsocthanksitem I. W. Tsang is with the Centre for Artificial Intelligence, University of Technology Sydney, Sydney 2007, Australia. \protect\\ E-mail: Ivor.Tsang@uts.edu.au.
\IEEEcompsocthanksitem H. Yin is with School of Information Technology and Electrical Engineering, The University of Queensland, Brisbane 4067, Australia. \protect\\ E-mail: h.yin1@uq.edu.au.
\IEEEcompsocthanksitem G. Yang and J. Li are
 with the School of Computer Science and Engineering, University of Electronic Science and Technology of China, Chengdu 611731, China. \protect\\  guowu@uestc.edu.cn, lijin117@yeah.net.
\IEEEcompsocthanksitem D. Lian is with School of Computer Science and Technology, University of Science and Technology of China, Hefei 230027, China. \protect\\ E-mail: dove.ustc@gmail.com.

}
}

\IEEEtitleabstractindextext{%
\begin{abstract}
Recommendation efficiency and data sparsity problems have been regarded as two challenges of improving performance for online recommendation. Most of the previous related work focus on improving recommendation accuracy instead of efficiency. In this paper, we propose a Deep Pairwise Hashing (DPH) to map users and items to binary vectors in Hamming space, where a user's preference for an item can be efficiently calculated by Hamming distance, which significantly improves the efficiency of online recommendation. To alleviate data sparsity and cold-start problems, the user-item interactive information and item content information are unified to learn effective representations of items and users. Specifically, we first pre-train robust item representation from item content data by a Denoising Auto-encoder instead of other deterministic deep learning frameworks; then we finetune the entire framework by adding a pairwise loss objective with discrete constraints; moreover, DPH aims to minimize a pairwise ranking loss that is consistent with the ultimate goal of recommendation. Finally, we adopt the alternating optimization method to optimize the proposed model with discrete constraints. Extensive experiments on three different datasets show that DPH can significantly advance the state-of-the-art frameworks regarding data sparsity and item cold-start recommendation.
\end{abstract}

}

\maketitle

\IEEEdisplaynontitleabstractindextext

%
\IEEEpeerreviewmaketitle

\section{Introduction}\label{sec:introduction}

%
%
%
%
\IEEEPARstart{P}ersonlized recommender systems have been recognized as one of the most critical and effective approaches for alleviating information overload, and also it is a key factor for the success of various applications such as online E-commerce webs sites: Amazon, Netflix, Yelp, etc., online educational systems, and even online health systems. Typically, a recommender system recommends a particular user with a small set from the underlying pool of items that the user may be interested in. Most of the existing recommendation models were based on users' previous behavior data such as ratings, purchasing records, click actions, and watching records, like/dislike records, etc. They were known as collaborative filtering (CF).

CF-based recommender systems are proven to be very successful. They produce top-$k$ items that users may be interested in by exploiting the historical interaction data. Among all CF-based methods, the latent factor models (e.g., matrix factorization) have been demonstrated to achieve great success in both academia and industry. Such CF-based methods factorize an user-item interaction matrix into a low-dimensional real latent space where both users and items are represented by real-valued vectors. Then the user's preference scores for items were predicted by inner products between real latent vectors, and the user's top-$k$ preferred items can be produced by ranking the scores descendingly, and this procedure is named online recommendation.

However, datasets in the real world are generally sparse, which leads to poor performance with CF-based recommender systems. Besides, CF-based recommender systems are incapable of handling cold-start problems~\cite{schein2002methods}, since new users or new items are lack of interaction data in cold-start settings. Another type of recommender system, the content-based recommendation, can offer suggestions based on the content similarities of users and items. To tackle the data sparsity and cold-start problems, several state-of-the-art content-aware recommender systems~\cite{wang2011collaborative,wang2015collaborative,zhang2016collaborative,wu2016collaborative} were proposed by combining the CF-based information and content-based information. Specifically, these content-aware recommender system extracted vector representations of new users or new items from their content data, and then these vector representations were unified into a CF-based framework, and thus it can be applied to cope with data sparsity and cold-start problems. 

In the previous content-aware recommender systems, they first learned real-valued representations for users and items, respectively, and then conducted a recommendation by an online similarity search procedure, online recommendation. A successful recommender system should meet the users' requirement of fast response to recommend items from a large set by analyzing their browsing, purchasing, and searching history. However, the growing scale of users and items has made the online recommendation much more challenging since the time cost with real-valued representations is expensive. 
Hence a few recommendation frameworks~\cite{yin2014lcars,yin2016adapting} were proposed to speed up the online recommendation. However, the time complexity has not yet been decreased markedly since these recommendations are still based on similarity search in the real space.

To speed up online recommendation fundamentally, several scholars put forward hashing-based recommendation frameworks~\cite{zhou2012learning,zhang2016discrete,zhang2017discrete} that can fundamentally improve the similarity search efficiency. To be specific, the hashing-based recommendation framework encodes users and items into binary codes in a Hamming space. Then the preference score, in this case, can be efficiently computed by bit operations, i.e., Hamming distance. One can even use a fast and accurate indexing method to find approximate top-\emph{k} preferred items with sublinear or logarithmic time complexity~\cite{wang2012semi}. So the time cost with hashing-based frameworks is significantly reduced compared with the real-based frameworks. Besides, each dimension of hash codes can be stored by only one bit instead of 32/64 bits that are used for storing one dimension of real-valued vectors, which significantly reduces storage cost. 
Most of the previous hashing-based recommender systems focus on rating prediction, 
but the rating-based objectives are not consistent with the ultimate goal of the recommender system -- providing a ranking list of items. Thus, a ranking-based objective should be more appropriate to formulate the recommendation task. Previous ranking-based recommendation such as Bayesian personalized ranking (BPR)~\cite{rendle2009bpr}, Cofi rank-maximum margin matrix factorization (CofiRank)~\cite{weimer2008cofi}, and List-wise learning to rank with matrix factorization (ListCF)~\cite{shi2010list}, have been proposed, showing superior performance of recommendation to rating-based methods. To speed up the online recommendation, Discrete Personalized Ranking (DPR)~\cite{zhang2017discrete}, were proposed based on a pair-wise ranking objective with discrete constraints. However, they cannot cope with data sparsity and cold-start issues.

To this end, this paper proposes a ranking-based hashing framework, Deep Pairwise Hashing (DPH), to alleviate data sparsity and cold-start problems, and also provide an efficient online recommendation. We have already presented our preliminary study of solving data sparsity and cold-start problems with an efficient way in Discrete Deep Learning (DDL)~\cite{zhang2018discrete}. This paper extendes DDL with an in-depth study and performance analysis. Specifically, this paper makes the following new contributions: first, to make the proposed recommendation framework applicable in a wide range of platforms, we design a new recommendation framework based on implicit feedbacks, since they are much more common in real world applications than explicit ratings; second, we design a ranking-based objective in the proposed DPH, that is consistent with the ultimate goal of recommendation -- providing a ranked items list for each user; third, to learn item robust representation, we choose Denoising Auto-encoder (DAE) instead of Deep Belief Network (DBN) embedded in DDL; and fourth, we conduct more extensive experiments to evaluate the performance of DDL and the enhanced DPH to overcome the cold-start and sparse problems on Amazon and Yelp datasets.
The main contributions of this paper are summarized as follows:
\begin{enumerate}
	\item We propose a pairwise ranking based hashing recommendation framework, which is capable of handling cold-start item and data sparsity issues, and providing efficient online recommendation. 
	\item We choose the Denoising Auto-encoder instead of other deterministic deep learning frameworks to learn robust item representation by modeling the noise, which is beneficial to improve the recommendation performance in sparse and item cold-start setting.


  \item We develop an alternating optimization algorithm by adding balance and irrelevant constraints on hash codes to solve the proposed discrete problem, which is helpful to extract compact and informative hash codes.
  
\end{enumerate}
The rest of this paper is organized as follows: Section~\ref{sec2} introduces two types of related work, the content-aware recommendation, and the hashing-based recommendation. Followed by the introduction of main notations appeared in this paper and the problem formulation in Section~\ref{sec3}. Next, we present the DPH framework and explain the detailed derivation of each step in Section~\ref{sec4}, and show the initialization and the discrete optimization algorithm of solving the proposed model in Section~\ref{sec5}. Explicitly, we first initialize DPH by unsupervised learning of DAE, and the robust item representations obtained are forwarded into the supervised DPH learning procedure. To optimize DPH, we adopt an alternating optimization algorithm composed of a series of mixed integer programming problems. Then, we introduce the experimental settings, description of datasets, and competing baselines in Section~\ref{sec6}, followed by the experimental analysis in Section~\ref{sec7}. Finally, we conclude this paper and disclose several future works could be done in Section~\ref{sec8}.



%
\section{Related Work}\label{sec2}
In this section, we review several major schemes closely relevant to this paper, that contains the content-aware recommendation, the ranking-based recommendation, and the hashing-based recommendation. But the three types have some overlaps, so we introduce the related works from the following two aspects: we first introduce several state-of-the-art content-aware recommender systems proposed to alleviate data sparsity and cold-start problems, we then introduce the latest two competing hashing-based recommendation frameworks. 
\subsection{Content-aware Recommender Systems}\label{sec2-a}
Collaborative topic regression~\cite{wang2011collaborative} is a state-of-the-art content-aware recommender system, which was proposed by combining the topic model, collaborative filtering, and probabilistic matrix factorization (PMF)~\cite{salakhutdinov2007probabilistic}. The authors developed a machine learning algorithm for recommending scientific articles to users in an online scientific community. CTR obtained real latent representations of users and items by exploiting two types of data: user's collection data and article content data. It can be used to mitigate cold start and data sparsity settings.

Another type of content-aware recommender systems is deep learning based framework \cite{zhang2019deep}. Collaborative deep learning (CDL)~\cite{wang2015collaborative} was proposed as a probabilistic model by jointly learning a probabilistic stacked denoising auto-encoder (SDAE)~\cite{vincent2010stacked} and CF. Similar to CTR, CDL exploits the interaction and content data to alleviate cold start and data sparsity problems. Differ from CTR, CDL took advantage of deep learning framework to learn effective real latent representations. Thus it can be applied in cold-start and sparse settings. CDL is also a tightly coupled method for recommender systems by developing a hierarchical Bayesian model.


Visual Bayesian Personalized Ranking~\cite{he2016vbpr} is a factorization model by incorporating visual features into predictors of users' preferences. By utilizing visual features extracted from product images by (\emph{pre-trained}) deep networks, VBPR is also helpful to alleviate cold start and sparse issues.

\subsection{Hashing-based Recommender Systems}\label{sec2-b}
A pioneer work~\cite{das2007google} was proposed to exploit Locality-Sensitive Hashing~\cite{datar2004locality} to generate hash codes for Google News readers based on their click history. On this basis, A. Karatzoglou et al.~\cite{karatzoglou2010collaborative} randomly projected real latent representations learned from regularized matrix factorization into hash codes. Similar to this, K. Zhou et al.\cite{zhou2012learning} followed the idea of Iterative Quantization~\cite{gong2013iterative} to generate binary codes from rotated real latent representations. To derive compact binary codes, the uncorrelated constraints were imposed on the real latent representations in regularized matrix factorization. However, according to the analysis in~\cite{zhang2014preference}, hashing essentially only preserves similarity rather than preference based on inner product, since the magnitudes of representations for users and items are discarded in the quantization stage. Thus, a constant feature norm (CFN) constraint was imposed when learning the real latent representations, and then the magnitudes and similarity are respectively quantized in~\cite{zhang2017dot}. But the two-stage approach still suffered large information loss in the quantization procedure. 

Differ from two-stage hashing frameworks, hashing learning frameworks can obtain hash codes by directly solving the discrete optimization problem. Thus more information is carried by hash codes than two-stage frameworks. Zhang et al. proposed discrete collaborative filtering (DCF)~\cite{zhang2016discrete}, which is a hashing learning recommendation framework. By adding balance and uncorrelated constraints on hash codes, DCF obtained efficient binary codes. DCF was evaluated using a similar way of the conventional CF~\cite{rendle2009bpr}. Then, a ranking-based hashing framework was proposed to improve the recommendation performance~\cite{zhang2017discrete} by adding the same constraints with DCF, and it also obtained short and informative hash codes. Since the above hashing learning frameworks are based on CF, thus they still suffer low recommendation accuracy under sparse setting, and they cannot work when new users or items present.
\section{Preliminary }\label{sec3}
\subsection{Notations}\label{sec3-1}
Let user and item index sets are denoted by $U=\{1,\cdots ,n\}$ and $I=\{1\cdots ,m\}$, respectively. $\mathbf{R}=(s_{ui})_{n\times m}$ is the observed implicit feedback matrix, thus entries in $\mathbf{R}$ contain only values `1' and `0', and $s_{ui} = 1 $ denotes user $u$ had an interaction with item $i$; otherwise there is no interactions between user $u$ and item $i$. Examples of user-item interactions contain users' purchases, clicks, collections, etc. Note that although we evaluate DPH on implicit feedbacks, but it also works on explicit ratings by transferring explicit ratings into implicit feedbacks. The index set of observed implicit feedbacks in $\mathbf{R}$ is denoted as $S$, which is a subset of the Cartesian product of $U$ and $I$: $S\subseteq U\times I$. Item content data is denoted as $\mathbf{C}$, which consists of bag-of-words vectors of all items. Other essential notations used in this paper are listed in Table~\ref{tab:1}.

\begin{table}
	\caption{Notations}
	\centering
	\begin{tabular}{ll}
		\toprule[1pt]
		Symbol  & Description \\
		\midrule[1pt]
		$\mathbf{B}$  & binary matrix of all users in $U$ \\
		$\mathbf{D}$  & binary matrix of all items in $I$ \\
		${{\mathbf{b}}_{u}}$  & hash code of the user $u$  \\
		${{\mathbf{d}}_{i}}$  & hash code of the item $i$  \\
		$I_{u}^{+}$  & items rated by the user $u$\\
		$I_{u}^{-}$  & items not rated by the user $u$\\
		$U_{i}^{+}$  & users rated the item $i$ \\ 
		$U_{i}^{-}$  & users not rated the item $i$ \\ 
		$S_{D}$  & the 2--tuple index set of observed interactions \\
		$S_{T}$  & the 3--tuple index set of observed interactions   \\
		$\Theta$ & parameters of SDAE \\
		$\mathbf{X}$  & the delegated real matrix of $\mathbf{B}$ \\
		$\mathbf{Y}$  & the delegated real matrix of $\mathbf{D}$ \\
		\bottomrule[1pt]
	\end{tabular}
	\label{tab:1}
\end{table}

\subsection{Denoising Auto-encoder}\label{sec3-2}
\begin{figure}
	\centering
	\includegraphics[width=0.8\linewidth]{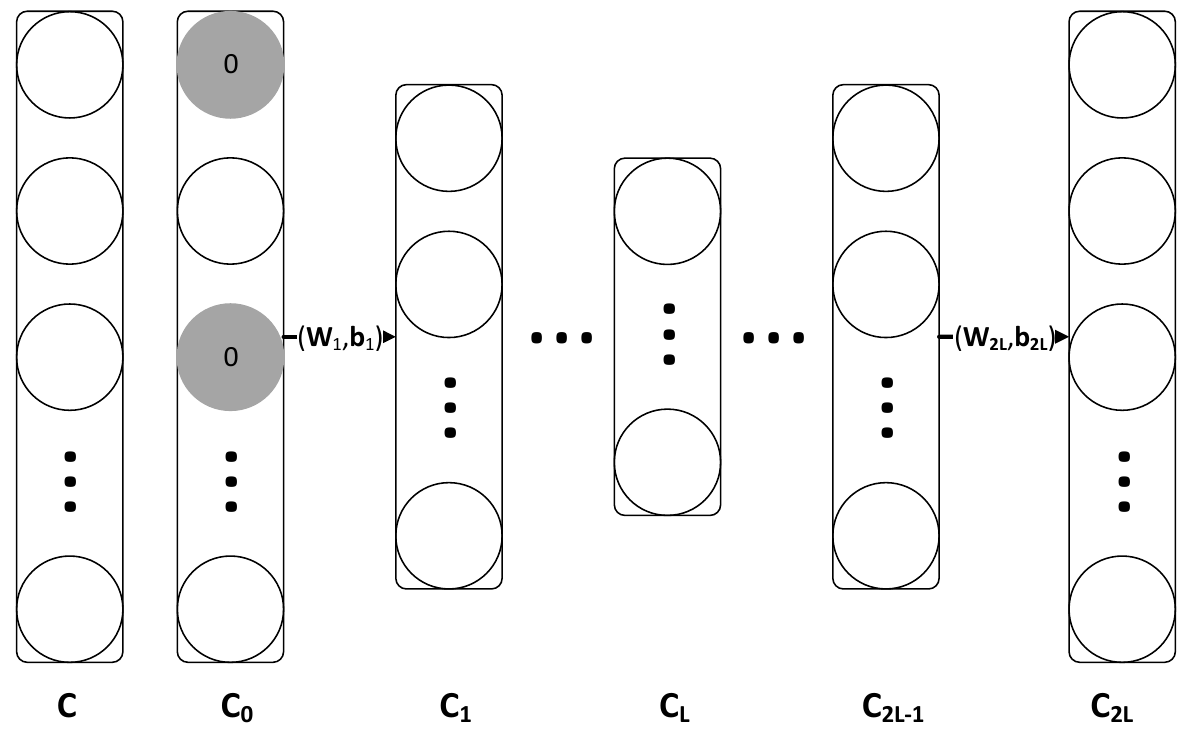} 
	\caption{2$L$-layer Denosing Auto-encoder.}\label{dae}
\end{figure}

The principle behind denoising auto-encoders is to be able to reconstruct data from an input of corrupted data. After giving the auto-encoder the corrupted data, we force the hidden layer to learn only the more robust representations. The output will then be a more refined version of the input data~\cite{vincent2008extracting}. Denoising Auto-encoders solve this problem by corrupting the data on purpose by randomly turning some of the input values to zero. Figure~\ref{dae} is a example of DAE with $2L$ layers. In general, the percentage of input nodes which are being set to zero depends on the amount of data and input nodes we have. Commonly the hidden layer in the middle. $\mathbf{C}_{L}$, is the latent representation we need, and the input layer $\mathbf{C}_{0}$ is the corrupted version of the clean input data $\mathbf{C}$. DAE solves the following optimization problem:
\begin{equation}
\underset{\mathbf{W}, \mathbf{b} }{\mathop{\arg \min }}\,{{{\left\| \mathbf{C}-\mathbf{C}_{L}\right\|}_{F}^{2}}} + \delta {\left\|\mathbf{W} \right\|}_{F}^{2}, \label{eq:dae}
\end{equation}
where $\Theta =\{\mathbf{W},\mathbf{b}\}$ is the parameters of $2L$-layer DAE that contains weight matrices $\mathbf{W}=\{\mathbf{W}_1, \mathbf{W}_2, \cdots, \mathbf{W}_{2L-1},\mathbf{W}_{2L} \}$ and bias vectors $\mathbf{b} = \{\mathbf{b}_1,\mathbf{b}_2, \cdots, \mathbf{b}_{2L-1}, \mathbf{b}_{2L} \}$, and $\delta $ is the hyper-parameter of the regularizer for generalization. 
\subsection{Problem Formulation}\label{sec3-3}
%

\textbf{Deep Pairwise Hashing:} Given a user-item interaction dataset $\mathbf{R}$, a content dataset $\mathbf{C}$, and an active user $u$, our goal is to recommend top-$k$ items that user $u$ would be interested in by a pairwise ranking function - Area under the ROC curve~\cite{bradley1997use} (AUC). The problem becomes a cold-start item recommendation when the predicted top-$k$ items are not in the interaction dataset $\mathbf{R}$, and meanwhile in the content data $\mathbf{C}$. 

\section{Deep Pairwise Hashing}\label{sec4}
\begin{figure*}
	\centering
	\includegraphics[width=1\linewidth]{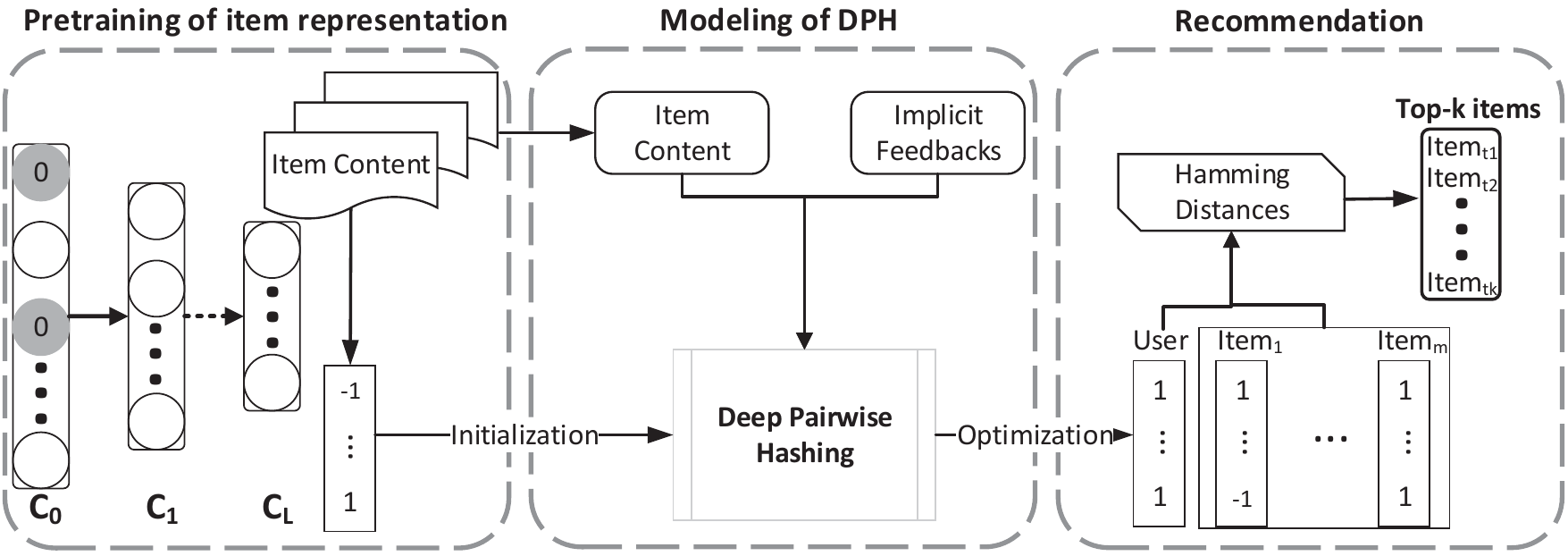} 
	\centering
	\caption{The DPH framework: we first obtain items' continuous embeddings with Denoising Auto-encoder by the left pretraining step from items' content data; we then integrate the obtained items' embeddings with implicit feedbacks to formulate the proposed DPH model with a ranking-based objective; we finally develop an alternating optimization method to solve the proposed objective and conduct recommendation with the learned hash codes in Hamming space. }\label{dph}
\end{figure*}

Deep Pairwise Hashing is proposed to deal with the efficient online recommendation and data sparsity by hashing technique and a content-aware objective, and the entire framework is shown in Figure~\ref{dph}. It initializes item representations by a denoising auto-encoder, and then obtains hash codes from a joint training of the content-aware objective and pairwise hashing objective, and finally conduct a recommendation for a specific user based on hash codes obtained.   
\subsection{Pairwise Ranking Objective}\label{sec4-1}
In this section, we first formulate user's preference by hash codes in Hamming Space. Followed by the traditional Matrix Factorization~\cite{koren2011advances}, we also formulate the preference as the similarity between representations of users and items. In Real Space, similarity can be defined as different metrics, such as cosine similarity, inner product, Euclidean distance, etc. While in Hamming Space, the most common similarity is the Hamming distance. Suppose users and items are presented as $r$-dimension hash codes, ${{\mathbf{b}}_{u}}\in {{\{\pm 1\}}^{r}}$ and ${{\mathbf{d}}_{i}}\in {{\{\pm 1\}}^{r}}$, respectively, and the Hamming distance $\text{H}(\mathbf{b}_{u}, \mathbf{d}_{i}) = \sum\limits_{k=1}^{r}{\mathbb{I}\left( {{b}_{uk}}\neq{{d}_{ik}} \right)}$, where $\mathbb{I}(\cdot)$ is an indicator function that returns 1 if the input is true, otherwise it returns 0. Then the preference of user $u$ for item $i$ is denoted by a expression of Hamming distance 
\label{key}\begin{equation}\label{eq:pre}
{\widehat{p}}_{ui}=1-\frac{1}{r}\text{H}(\mathbf{b}_{u}, \mathbf{d}_{i})=\frac{1}{2}+\frac{1}{2r}\mathbf{b}_{u}^{T}{{\mathbf{d}}_{i}}
\end{equation}
where ${\widehat{p}}_{ui}$ is within the range of 0 to 1, that represents the predicted preference of the user $u$ over the item $i$. From the above preference predicted equation, we observe that preference metric in the $r$-dimension Hamming space is consistent with the $r$-dimension Real space where preference is predicted by inner products of real latent vectors.

Discrete Pairwise Hashing (DPH) is devised as a pairwise framework for top-$k$ recommendation. For a specific user $u$, let the hash code of user $u$ be $\mathbf{b}_u$, and hash codes of item $i$ and item $j$ be $\mathbf{d}_i$ and $\mathbf{d}_j$, respectively, then the predicted pairwise preference on item $i$ and item $j$ are denoted as ${\widehat{p}}_{uij}$,
\begin{equation}
{{\widehat{p}}_{uij}}={{\widehat{p}}_{ui}}-{{\widehat{p}}_{uj}}.
\end{equation}
If ${{\widehat{p}}_{uij}}>0$, we consider that user $u$ prefers item $i$ over $j$; otherwise, user $u$ prefers item $j$ over $i$. For implicit feedbacks, positive feedbacks refer to those `1' entries. Specifically, for the user $u$, the positive items are those rated by the user, and the negative ones are those not rated. We expect the predicted preference of the positive item $i$, $\widehat{p}_{ui}$, would be greater than the negative item $j$, $\widehat{p}_{uj}$, i.e., $\widehat{p}_{uij} >0$. If we regard the pairwise ranking task as a two-class classification problem, then we choose a common metric, the Area under the ROC (AUC), as the objective. According to~\cite{rendle2009bpr}, AUC of each user $u$ is defined as
\begin{equation}
\text{AUC}(u)=\frac{1}{\left| I_{u}^{+} \right|\cdot \left| I_{u}^{-} \right|}\sum\limits_{i\in I_{u}^{+}}{\sum\limits_{j\in I_{u}^{-}}{\mathbb{I} \left( {{{\widehat{p}}}_{uij}}>0 \right)}},
\end{equation}
where $ I_{u}^{+}$ and $I_{u}^{-}$ are defined in Table~\ref{tab:1}. The value of $\text{AUC}(u)$ presents the average `correct' classification pairwise samples. The `correct' classification means positive items have greater predicted preference value than negative items. $\text{AUC}(u)$ ranges from zero to one. Two extreme cases are: $\text{AUC}(u)=1$ means all preferences are preserved, and the learned representations of users and items are perfect to predict preferences; $\text{AUC}(u)=0$ means the no preferences are preserved with the learned representations. 
\begin{equation}
\text{AUC}=\frac{1}{\left| U \right|}\sum\limits_{u\in U}{\text{AUC}\left( u \right)}
\end{equation}
To simplify the following expressions, we define ${S}$ as ${S}=\{\left( u,i,j \right)\left| u\in U,i\in  \right.I_{u}^{+}\ \text{and}\ j\in I_{u}^{-}\}$, and denote the coefficient $z_u = \frac{1}{\left| U \right| \left| I_{u}^{+} \right| \left| I_{u}^{-} \right|}$. Then we rewrite AUC as
\begin{equation}
\text{AUC} =\sum\limits_{\left( u,i,j \right)\in {S}}z_u \mathbb{I}\left( {{{\widehat{p}}}_{ui}}>{{{\widehat{p}}}_{uj}} \right)
\end{equation}
However, the above objective is a discrete problem, thus optimizing AUC directly often leads to an NP-hard problem~\cite{gao2013one}. A feasible solution in practice is to minimize some pairwise surrogate losses
\begin{equation}
L = \sum\limits_{\left( u,i,j \right)\in {S}}z_u \ell({{{\widehat{p}}}_{ui}} - {{{\widehat{p}}}_{uj}})
\end{equation}
where $\ell: \mathbb{R} \rightarrow \mathbb{R}^+$ is a convex function such as exponential loss $\ell(t) = e
^{-t}$, hinge loss $\ell(t)=\max(0, 1-t)$, logistic loss $\ell(t) = \log(1 + e^{-t})$, least square loss $\ell(t) = (1-t)^2$, etc.

As the least square loss is consistent with AUC~\cite{gao2013one} and could lead to efficient and closed forms for updating real latent vectors without sampling in an either continuous or discrete case. So we propose to use the least square loss $\ell(t) = (1-t)^2$, and to minimize the following pairwise least square loss:
\begin{equation}\label{eq:objective}
\min \sum\limits_{\left( u,i,j \right)\in {S}}z_u{{\left( 1-\left( {{{\widehat{p}}}_{ui}}-{{{\widehat{p}}}_{uj}} \right) \right)}^{2}},
\end{equation}
In this paper, we are interested in mapping users and items into $r$-dimension binary codes for fast recommendation, where user-item similarity can be efficiently calculated via Hamming distance in the $r$-dimension Hamming space. If we stack users' hash codes $\mathbf{b}_u \in \left\{ \pm 1 \right\}^r$ and items' hash codes $\mathbf{d}_i \in \left\{ \pm 1 \right\}^r$ by column into matrix $\mathbf{B}\in {{\left\{ \pm 1 \right\}}^{r\times n}}$ and $\mathbf{D}\in {{\left\{ \pm 1 \right\}}^{r\times m}}$, respectively. Then, by substituting Equation~\eqref{eq:pre} into Equation~\eqref{eq:objective}, we derive the Pairwise Hashing objective as follows:
\begin{equation}\label{eq:hashing}
\underset{\mathbf{B},\mathbf{D}}{\mathop{\arg \min }}\,\sum\limits_{\left( u,i,j \right)\in {S}}{z_u{{\left( 2r-\mathbf{b}_{u}^{T}\left( {{\mathbf{d}}_{i}}-{{\mathbf{d}}_{j}} \right) \right)}^{2}}}
\end{equation}

\subsection{Binary Content-aware Objective}\label{sec4-2}

To solve data sparsity and cold-start item problems, we incorporate content and interaction data into the proposed DPH framework. In this section, we introduce how to learn item robust representation from content data.

Deep learning models have been widely demonstrated as the most successful representation learning. 
Besides, data in our real world often contains some noise, and we know that DAE can be applied for extracting robust representation from Section~\ref{sec3-2}, so we choose DAE for learning cold-start item representation in this paper.

For each item $i$, suppose the bag-of-words vector from item content data to be $\mathbf{c}_i$, we first extract real latent representation $\mathbf{f}_{i}$ from the middle layer (the $L$-th layer) by training a standard $2L$-layer DAE based on Equation~\eqref{eq:dae} after randomly turning some of the input values to zero, and we denote the real latent representation $\mathbf{f}_{i}$ as
\begin{equation}
{{\mathbf{f}}_{i}}=\text{DAE}({\mathbf{c}_{L,i}},\Theta ).
\end{equation}
To derive hash codes, we directly minimize the
difference between the real latent representation ${\mathbf{f}_{i}}$ and the expected hash code ${\mathbf{d}_{i}}$, and it is given by
\begin{equation}\label{eq:content}
\underset{\mathbf{D},\ \Theta }{\mathop{\arg \min }}\,\sum\limits_{i=1}^{m}{{{\left\| {{\mathbf{d}}_{i}}-\text{DAE}({{\mathbf{c}}_{L,i}},\Theta ) \right\|}_{F}^{2}}},
\end{equation}
where $\Theta =\{\mathbf{W},\mathbf{b}\}$ is parameters of the $2L$-layer DAE that contains weight matrices $\mathbf{W}=\{{{\mathbf{W}}_1},{{\mathbf{W}}_2},\cdots ,{\mathbf{W}}_{2L}\}$ and bias vectors $\mathbf{b}=\{{{\mathbf{b}}_1},{{\mathbf{b}}_2},\cdots ,{{\mathbf{b}}_{2L}}\}$. $\mathbf{D}$ is stacked by $\mathbf{d}_{i}$, where $i \in I$. By minimizing the objective in Equation~\eqref{eq:content}, we obtain effective item hash code from content data.

\subsection{Deep Pairwise Hashing}
To improve recommendation accuracy and efficiency, we combine a pairwise ranking based hashing objective~\eqref{eq:hashing} and a content-aware objective~\eqref{eq:content} as the objective of Deep Pairwise Hashing 
\begin{align}\label{eq:obj3}
  \underset{\mathbf{B,D,}\Theta }{\mathop{\arg \min }}\, &\sum\limits_{\left( u,i,j \right)\in {S}}{z_u{{\left( 2r-\mathbf{b}_{u}^{T}\left( {{\mathbf{d}}_{i}}-{{\mathbf{d}}_{j}} \right) \right)}^{2}}} \notag\\
  & +\lambda\sum\limits_{i=1}^{m}{{{\left\| {{\mathbf{d}}_{i}}-\text{DAE}({{\mathbf{c}}_{L,i}},\Theta) \right\|}_{F}^{2}}}\notag\\
  s.t.\ \ & \mathbf{B}\in {{\left\{ \pm 1 \right\}}^{r\times n}},\mathbf{D}\in {{\left\{ \pm 1 \right\}}^{r\times m}},
\end{align}
where $\lambda>0$ is a hyper parameter that weights the importance of two objectives. To maximize the entropy of each binary bit, we add a balance constraint, so that each bit carries as much information as possible. In addition, to learn compact binary codes, we impose irrelevant constraints on hash codes, thus, we can obtain hash codes with no redundant information. Then the above problem in Equation~\eqref{eq:obj3} is reformulated as
\begin{align}\label{eq:obj4}
  \underset{\mathbf{B,D,}\Theta }{\mathop{\arg \min }}\, &\sum\limits_{\left( u,i,j \right)\in {S}}{z_u{{\left( 2r-\mathbf{b}_{u}^{T}\left( {{\mathbf{d}}_{i}}-{{\mathbf{d}}_{j}} \right) \right)}^{2}}} \notag \\
   &+\lambda\sum\limits_{i=1}^{m}{{{\left\| {{\mathbf{d}}_{i}}-\text{DAE}({{\mathbf{c}}_{L,i}},\Theta ) \right\|}_{F}^{2}}} \notag \\
  s.t. \ \ & \mathbf{B}\in {{\left\{ \pm 1 \right\}}^{r\times n}},\mathbf{D}\in {{\left\{ \pm 1 \right\}}^{r\times m}}, \notag\\
 &\mathbf{B1}_{n}=\mathbf{0},\mathbf{D1}_{m}=\mathbf{0},\mathbf{B}{{\mathbf{B}}^{T}}\mathbf{=}n{{\mathbf{I}}_{r}}\mathbf{,D}{{\mathbf{D}}^{T}}\mathbf{=}m{{\mathbf{I}}_{r}},
\end{align}
where $\mathbf{1}_{n}$ ($\mathbf{1}_{m}$) are  $n$-dimension ($m$-dimension) all `1' entries vectors, and $\mathbf{I}_{r}$ is a $r\times r$-dimension identity matrix. As the problem~\eqref{eq:obj4} is a discrete optimization problem, which is proven to be an intractable NP-hard problem~\cite{haastad2001some}, we adopt a methodology like~\cite{zhang2016discrete} to soften the balance and irrelevant constraints. Specifically, we add a delegated real valued $r\times n$-dimension matrix $\mathbf{X}$ and a $r\times m$-dimension $\mathbf{Y}$ to approximate hash codes $\mathbf{B}$ and $\mathbf{D}$, respectively. Thus the problem~\eqref{eq:obj4} can be rewritten as:
\begin{align}\label{eq:obj05}
\underset{\mathbf{B,D,}\Theta }{\mathop{\arg \min }}\, &\sum\limits_{\left( u,i,j \right)\in {S}}{z_u{{\left( 2d-\mathbf{b}_{u}^{T}\left( {{\mathbf{d}}_{i}}-{{\mathbf{d}}_{j}} \right) \right)}^{2}}} + \alpha \left\| \mathbf{B}-\mathbf{X} \right\|_{F}^{2}\notag\\
 &+\lambda \sum\limits_{i=1}^{m}{\left\| {{\mathbf{d}}_{i}}-\text{DAE}({{\mathbf{c}}_{L,i}},\Theta ) \right\|_{F}^{2}}+\beta \left\| \mathbf{D}-\mathbf{Y} \right\|_{F}^{2}, \notag\\
 &  s.t.\ \ \ \mathbf{B}\in {{\left\{ \pm 1 \right\}}^{r\times n}},\mathbf{D}\in {{\left\{ \pm 1 \right\}}^{r\times m}} \notag\\
  &\mathbf{X}{{\mathbf{1}}_{n}}=\mathbf{0},\mathbf{Y}{{\mathbf{1}}_{m}}=\mathbf{0},\mathbf{X}{{\mathbf{X}}^{T}}\mathbf{=}n{{\mathbf{I}}_{r}},\mathbf{Y}{{\mathbf{Y}}^{T}}\mathbf{=}m{{\mathbf{I}}_{r}},
\end{align}
where $\alpha$ and $\beta$ are hyper parameters that allows certain discrepancy between $\mathbf{B}$ and $\mathbf{X}$, and between $\mathbf{D}$ and $\mathbf{Y}$. Since $\text{\emph{tr}}(\mathbf{B}\mathbf{B}^T) = \text{\emph{tr}}(\mathbf{X}\mathbf{X}^T)=nr$ and $\text{\emph{tr}}(\mathbf{D}\mathbf{D}^T) = \text{\emph{tr}}(\mathbf{Y}\mathbf{Y}^T)=mr$ are constant. Thus the objective in Equation~\eqref{eq:obj05} can be equivalently transformed as the following mixed integer optimization problem:
\begin{align}\label{eq:obj5}
  \underset{\mathbf{B,D,}\Theta ,\mathbf{X,Y}}{\mathop{\arg \min }}\,&\sum\limits_{\left( u,i,j \right)\in {S}}{z_u{{\left( 2r-\mathbf{b}_{u}^{T}\left( {{\mathbf{d}}_{i}}-{{\mathbf{d}}_{j}} \right) \right)}^{2}}} - 2\alpha tr({{\mathbf{B}}^{T}}\mathbf{X}) \notag \\
 & +\lambda\sum\limits_{i=1}^{m}{{{\left\| {{\mathbf{d}}_{i}}-\text{DAE}({{\mathbf{c}}_{L,i}},\Theta ) \right\|}_{F}^{2}}}-2\beta tr({{\mathbf{D}}^{T}}\mathbf{Y})\notag \\
 s.t.\ \ &\mathbf{B}\in {{\left\{ \pm 1 \right\}}^{r\times n}},\mathbf{D}\in {{\left\{ \pm 1 \right\}}^{r\times m}}\notag\\
 &\mathbf{X1}_n=\mathbf{0},\mathbf{Y1}_m=\mathbf{0},\mathbf{X}{{\mathbf{X}}^{T}}\mathbf{=}n{{\mathbf{I}}_{r}},\mathbf{Y}{{\mathbf{Y}}^{T}}\mathbf{=}m{{\mathbf{I}}_{r}}.
\end{align}
Differ from two-stage hashing-based frameworks, we do not discard the binary constraints through the objective transformations, which avoids  information loss caused by the quantization stage. By optimizing the above mixed integer problem, we obtain compact and informative hash codes from user-item interaction data and item content data with two constraints of balance and un-correlation. 
\subsection{Model Optimization}\label{sec5}
In this section, an alternating optimization method is developed to solve the mixed integer optimization problem shown in Equation~\eqref{eq:obj5}. We first initialize all parameters of DPH in Section~\ref{sec5-a}. We then introduce how to update $\mathbf{B}$, $\mathbf{D}$, $\Theta$, $\mathbf{X}$, and $\mathbf{Y}$, respectively in the following subsections.
\subsubsection{Initialization}\label{sec5-a}
To learn effectiveness cold-start item representation, we first apply DAE on the corrupted bag-of-words vector to get robust real latent representation, and then we initialize item's hash code $\mathbf{d}_i = sgn(\text{DAE}(\mathbf{c}_{L,i},\Theta))$, where the function $sgn(x)$ returns $1$ if $x>0$, and $-1$ otherwise. We call this initialization procedure as `Pretraining' in Figure~\ref{dph}. Similarly, we initialize $\Theta$ as the result of pretraining.
For other parameters, we initialize them by the Gaussian randomly strategy. Specifically, we initialize $\mathbf{X}$ and $\mathbf{Y}$ by the standard Gaussian distribution with mean $\mathbf{0}$ and variance $\mathbf{I}$, independently, and initialize $\mathbf{B}$ by the sign of $\mathbf{X}$. Next, we will introduce our alternating optimization method to learn hash codes of users and items, respectively.

\subsubsection{Update $\mathbf{B}$}\label{sec5-b}
Since the objective function in Equation~\eqref{eq:obj5} sums over users independently given $\mathbf{D}$, $\mathbf{X}$, $\mathbf{Y}$, and $\Theta$, we update $\mathbf{B}$ by updating each ${{\mathbf{b}}_{u}}$ in parallel by minimizing the following objective function:
\begin{multline}\label{eq:Bsub}
\underset{{{\mathbf{b}}_{u}}\in {{\left\{ \pm 1 \right\}}^{r}}}{\mathop{\arg \min }}\sum\limits_{(i,j)\in{S}_{u}}z_{u}\left( ( {{\left( {{\mathbf{d}}_{i}}-{{\mathbf{d}}_{j}} \right)}^{T}}{{\mathbf{b}}_{u}} )^{2} \right. \\
\left. \ -4r{{\left( {{\mathbf{d}}_{i}}-{{\mathbf{d}}_{j}} \right)}^{T}}{{\mathbf{b}}_{u}} \right)-2\alpha {\mathbf{x}}_{u}^{T}{{\mathbf{b}}_{u}},
\end{multline}
where $S_u = \{(i,j)|i\in I_{u}^{+}, j\in I_{u}^{-} \}$. This discrete optimization subproblem is NP-hard, and thus we adopt a bitwise learning strategy named Discrete Coordinate Descent (DCD)~\cite{Shen_2015_CVPR} to update ${{\mathbf{b}}_{u}}$. Particularly, let ${{b}_{uk}}$ be the $k$-th bit of ${{\mathbf{b}}_{u}}$ and let ${{\mathbf{b}}_{u\bar{k}}}$ be the rest bits of ${{\mathbf{b}}_{u}}$, i.e, ${{\mathbf{b}}_{u}}={{[\mathbf{b}_{u\bar{k}}^{T},{{b}_{uk}}]}^{T}}$. Note that DCD can update ${{b}_{uk}}$ given ${{\mathbf{b}}_{u\bar{k}}}$. Discarding terms independent of ${{b}_{uk}}$, the objective function in Equation~\eqref{eq:Bsub} is rewritten as
\begin{equation}\label{eq:Bsub1}
\underset{{{b}_{uk}}\in \{\pm 1\}}{\mathop{\arg \min }}\,{{\hat{b}}_{uk}}{{b}_{uk}},
\end{equation}
where ${{{\hat{b}}}_{uk}}=\sum\limits_{i,j\in S_u}z_{u}((\mathbf{d}_{i\bar{k}}-\mathbf{d}_{j\bar{k}})^{T}\mathbf{b}_{u\bar{k}}(d_{ik}-d_{jk}) - 2r d_{ik}+2r d_{jk})-\alpha x_{uk}$. Due to the space limit, we omit the derivation details. The above objective function reaches the minimal only if ${{b}_{uk}}$ had the opposite sign of ${{\hat{b}}_{uk}}$. However, if ${{\hat{b}}_{uk}}$ was zero, ${{b}_{uk}}$ would not be updated. Therefore, the update rule of ${{b}_{uk}}$ is
\begin{equation}\label{eq:updateB}
{{b}_{uk}}= -sgn \left( K\left( {{{\hat{b}}}_{uk}},{{b}_{uk}} \right) \right),
\end{equation}
where 
\begin{equation}
K({\hat{b}}_{uk},{b}_{uk})=\left\{\begin{aligned}
& {\hat{b}}_{uk},\ \text{if}\ {\hat{b}}_{uk}\ne 0;  \\ 
& -{b}_{uk},\ \text{otherwise}. \\ 
\end{aligned} \right.
\end{equation}

\subsubsection{Update $\mathbf{D}$}\label{sec5-c}
Given $\mathbf{B}$, $\mathbf{X}$, $\mathbf{Y}$, and $\Theta $, we derive the following subproblem regarding ${{\mathbf{d}}_{i}}$ by discarding terms irrelevant to ${{\mathbf{d}}_{i}}$ in Equation~\eqref{eq:obj5}:
\begin{multline}\label{eq:Dsub}
\underset{{{\mathbf{d}}_{i}}\in {{\left\{ \pm 1 \right\}}^{r}}}{\mathop{\arg \min }}\sum_{(u,j)\in S_{u}^{+}}{z_{u}{{\left( 2r-{{\mathbf{b}}_{u}}^{T}\left( {{\mathbf{d}}_{i}}-{{\mathbf{d}}_{j}} \right) \right)}^{2}}}\  \\
+\sum_{(u,j)\in S_{u}^{-}}{z_{u}{{\left( 2r-{{\mathbf{b}}_{u}}^{T}\left( {{\mathbf{d}}_{j}}-{{\mathbf{d}}_{i}} \right) \right)}^{2}}} \\
+\lambda{{({{\mathbf{d}}_{i}}-\mathbf{f}_{i})}^{2}}-2\beta{{\mathbf{y}}_{i}}^{T}{{\mathbf{d}}_{i}},
\end{multline}
where ${S}_{i}^{+}=\{(u,j)|u\in U_{i}^{+}, j \in I_{u}^{-} \}$, ${S}_{i}^{-}=\{(u,j)|u\in U_{i}^{-}, j \in I_{u}^{+} \}$. Similarly, we optimize ${{\mathbf{d}}_{i}}$ by the bitwise learning, and obtain the following update rule:
\begin{equation}\label{eq:updateD}
{{d}_{ik}}= - sgn \left( K\left( {{{\hat{d}}}_{ik}},{{d}_{ik}} \right) \right),
\end{equation}
where 
\begin{multline}
{{{\hat{d}}}_{ik}}=
\sum\limits_{(u,j)\in S_{u}^{+}}z_{u}\left(-{{d}_{jk}}-2r{{b}_{uk}}+\mathbf{b}_{u\bar{k}}^{T}\left( {{\mathbf{d}}_{i\bar{k}}}-{{\mathbf{d}}_{j\bar{k}}} \right){{b}_{uk}}\right)\\
-\sum\limits_{(u,j)\in S_{u}^{-}}z_{u}\left(\mathbf{b}_{u\bar{k}}^{T}\left( {{\mathbf{d}}_{j\bar{k}}}-{{\mathbf{d}}_{i\bar{k}}}\right){{b}_{uk}}+{{d}_{jk}}-2r{{b}_{uk}} \right) \\
-\lambda{{f}_{ik}}-\beta {{y}_{ik}}.
\end{multline}
\subsubsection{Update $\Theta$}\label{sec5-d}
Given $\mathbf{B}$, $\mathbf{D}$, $\mathbf{X}$, and $\mathbf{Y}$, the optimization problem~\eqref{eq:obj5} is rewritten as
\begin{equation}\label{eq:Tsub}
\underset{\Theta }{\mathop{\arg \min }}\,\sum\limits_{i=1}^{m}{{{\left\| {{\mathbf{d}}_{i}}-\text{DAE}({{\mathbf{c}}_{L,i}},\Theta ) \right\|}^{2}}}.
\end{equation}
Problem~\eqref{eq:Tsub} becomes a supervised DAE learning task. We often use stochastic gradient descent method to fine-tune parameters of DAE, where the gradient descent part is implemented by the Back-propagation algorithm. As ${{\mathbf{d}}_{i}}\in {{\{\pm 1\}}^{r}}$, we choose \emph{tanh} function as the output function of DAE since its output is in the same range $[-1,1]$ with hash codes. We choose sigmoid function as activation functions of hidden layers.
\subsubsection{Update $\mathbf{X}$}\label{sec5-e}
Given $\mathbf{B}$, $\mathbf{D}$, $\mathbf{Y}$, and $\Theta$, the Equation~\eqref{eq:obj5} is transformed as
\begin{equation}\label{eq:Xsub}
\underset{\mathbf{X}\in {{\mathbb{R}}^{r\times n}}}{\mathop{\text{argmax}}}\,tr({{\mathbf{B}}^{T}}\mathbf{X}),\ s.t.\ \mathbf{X1}_n=\mathbf{0},\mathbf{X}{{\mathbf{X}}^{T}}\mathbf{=}n\mathbf{I}_r.
\end{equation}
It can be solved with the help of SVD according to~\cite{zhang2016discrete, zhang2017discrete}. Specifically, $\mathbf{X}$ is updated by
\begin{equation}\label{eq:updateX}
\mathbf{X}=\sqrt{n}[{{\mathbf{U}}_{s}}\ {{\widehat{\mathbf{U}}}_{s}}]{{[{{\mathbf{V}}_{s}}\ {{\widehat{\mathbf{V}}}_{s}}]}^{T}},
\end{equation}
where ${{\mathbf{U}}_{s}}$ and ${{\mathbf{V}}_{s}}$ are respectively stacked by the left and right singular vectors of the row-centered matrix $\mathbf{\bar{B}}: {{\bar{b}}_{ui}}={{b}_{ui}}-\frac{1}{n}\sum\limits_{u=1}^{n}{{{b}_{ui}}}$. ${{\widehat{\mathbf{U}}}_{s}}$ is stacked by the left singular vectors and ${{\widehat{\mathbf{V}}}_{s}}$ can be calculated by Gram-Schmidt orthogonalization, and it satisfies ${{[{{\mathbf{V}}_{s}}\ \mathbf{1}]}^{T}}{{\widehat{\mathbf{V}}}_{s}}=\mathbf{0}$.
\subsubsection{Update $\mathbf{Y}$}\label{sec5-f}
Given $\mathbf{B}$, $\mathbf{D}$, $\mathbf{X}$, and $\Theta$, the Equation~\eqref{eq:obj5} can be transformed as
\begin{equation}\label{eq:Ysub}
\underset{\mathbf{Y}\in {{\mathbb{R}}^{r\times m}}}{\mathop{\text{argmax}}}\,tr({{\mathbf{D}}^{T}}\mathbf{Y}),\ s.t.\ \mathbf{Y1_m}=\mathbf{0},\mathbf{Y}{{\mathbf{Y}}^{T}}\mathbf{=}m\mathbf{I}_r.
\end{equation}
Similar to Section~\ref{sec5-e}, $\mathbf{Y}$ can be updated by
\begin{equation}\label{eq:updateY}
\mathbf{Y}=\sqrt{m}[{{\mathbf{P}}_{s}}\ {{\widehat{\mathbf{P}}}_{s}}]{{[{{\mathbf{Q}}_{s}}\ {{\widehat{\mathbf{Q}}}_{s}}]}^{T}}.
\end{equation}
Similarly, ${\mathbf{P}}_{s}$, $\mathbf{Q}_{s}$, ${{\widehat{\mathbf{P}}}_{s}}$, and ${{\widehat{\mathbf{Q}}}_{s}}$ can be determined by the row-centered matrix of $\mathbf{D}$.
\subsubsection{Algorithm}
We integrate the above initialization and alternating optimization methods into Algorithm~\ref{alg:dph}. We choose the optimal hyper-parameters $\alpha$, $\beta$ and $\lambda$ by grid search from $[10^{-5}, 10^5]$ on validation datasets. Although, the algorithm composed of an outside loop of the alternating optimization on $\mathbf{B}$, $\mathbf{D}$, $\mathbf{X}$, $\mathbf{Y}$ and $\Theta$, and a inner loop of updating each $\mathbf{b}_{u}$ ($\mathbf{d}_{i}$). It costs about 50 times to convergence for the outside loop. For the inner loop, it costs about two or three times to convergence, so the training cost is acceptable.  
\begin{algorithm}
	\KwIn{User-item implicit feedback $\mathbf{R}$, item conent data $\mathbf{C}$, DAE layer structure $[8000, 200, 30]$, $\alpha, \beta, \lambda$;}
	\KwOut{User hash codes $\mathbf{B}$ and item hash codes $\mathbf{D}$;}
	\caption{\text{Deep Pairwise Hashing (DPH)}}\label{alg:dph} 
	Pretrain: $\Theta \leftarrow$ Equation~\eqref{dae}\;
	Initialize: $\mathbf{X}, \mathbf{Y}\leftarrow \mathcal{N}(\mathbf{0},\mathbf{I}),\ \mathbf{B}\leftarrow sgn(\mathbf{X}), \ \mathbf{D}\leftarrow sgn(\text{DAE}(\mathbf{c}_{L,i},\Theta))$ \;
	\Repeat{Equation~\eqref{eq:obj5} convergence}	
	{	
		\For{$u \in \{1,\cdots, n\}$}{
			\Repeat{$\mathbf{b}_{u}$ convergence} 
			{
			\For{$k \in \{1,\cdots,r\}$}
			  {
				update $b_{uk} \leftarrow$ Equation~\eqref{eq:updateB} \;
			  }
		    }
	    }
		\For{$i \in \{1,\cdots,m \}$}
		{
		\Repeat{$\mathbf{d}_{i}$ convergence}
	     	{
		\For{$k \in \{1,\cdots,r \}$}{
			update $d_{ik} \leftarrow$ Equation~\eqref{eq:updateD} \;
		        }
	        }
	    }
     update $\mathbf{X}\leftarrow$ Equation~\eqref{eq:updateX} \;
     update $\mathbf{Y}\leftarrow$ Equation~\eqref{eq:updateY} \;
     finetune $\Theta$ by adding the objective Equation~\eqref{eq:Tsub} \; 
    }
\end{algorithm}

\section{Experiments}\label{sec6}
We compare the proposed DPH with competing hashing baselines and content-aware real-valued baselines. Experiments show that DPH outperforms the competing baselines in various sparse and cold-start environments on Amazon and Yelp datasets. We also discuss the advantage of hashing based recommender systems over the real-valued frameworks, and show that hashing methods significantly improve the efficiency of online recommendation, and thus it is adaptable to the increasing items number with the E-commerce development.

\subsection{Datasets}\label{sec6-a}
\subsubsection{Data Description}
To verify the effectiveness of the proposed DPH in cold-start and data sparsity settings, we choose two high sparse public datasets with the sparsity level of 99.9\% (ratio of unobserved ratings to the users' size times the items' size), Amazon dataset\footnote[2]{http://jmcauley.ucsd.edu/data/amazon} and Yelp Challenge Round 9 dataset\footnote[3]{https://www.yelp.com/dataset/challenge}. Amazon dataset covers 142.8 million user-item interactions (ratings, review text, helpfulness votes), as well as item content (descriptions, category, etc.) on 24 product categories spanning May 1996 - July 2014. Yelp Round 9 dataset contains 4.1M reviews (ratings, reviews) and 947K tips by 1M users for 144K businesses in cites of UK and US. We separately choose two of the largest subsets: `Clothing, Shoes \& Jewelry' dataset from Amazon, and `Phoenix' dataset from Yelp in our experiments. 
The detailed statistics of the two rating datasets are displayed in Table~\ref{tab:2}.

\subsubsection{Data Pre-processing}

We first transfer explicit ratings into implicit feedbacks by setting all observed ratings to positive feedbacks `1' and all unobserved to negative feedbacks `0', and these implicit feedbacks are also called ratings in this paper. 

For content data, we first remove punctuations, numbers, stop words, and words with the length smaller than two since these words usually have no discriminative meanings, we then conduct stemming on the remaining words by the Porter Stemmer~\cite{porter1980algorithm}. Finally, similar to~\cite{wang2011collaborative}, we choose top 8000 discriminative words from two datasets to form dictionaries separately by sorting the TF-IDF~\cite{ramos2003using} values from high to low, then we get the bag-of-words $\mathbf{C}$ of all items.
\subsubsection{Data Spliting}
To verify the effectiveness of DPH in various extremely sparse and cold-start settings, we do experiments by adding another sparsity levels the above two rating datasets like~\cite{ma2008sorec}. For example, if we set the sparsity level 10\% on the Amazon rating dataset, that means we randomly select 10\% implicit ratings as training dataset, $D_{train}$, from the original Amazon dataset, and then the sparsity of $D_{train}$ raised to 99.997\%, and we finally test the recommendation performance of all methods on the remaining ratings, denoted as $D_{test}$. The random selection is carried out five times independently, and
we report the experimental results as the average values.

\begin{table}
  \caption{Statistics of datasets.}
  \centering
  \begin{tabular*}{\hsize}{@{}@{\extracolsep{\fill}}c|c|c|c|c@{}}
  \hline
  Dataset & \#User & \#Item & \#Positive Feedbacks& Sparsity(\%)\\
  \hline
   Amazon & 39,387 & 23,033 & 278,653 & 99.97\%\\
    \hline
 Yelp & 122,097 & 11,854 & 353,772 & 99.98\%  \\
  \hline
  \end{tabular*}
 \label{tab:2}
\end{table}
\subsection{Experimental Settings}
\subsubsection{Evaluation Methods}\label{sec6-c}
As introduced in Section~\ref{sec3-3}, the goal of recommendation is to find out the top-$k$ items that users may be interested in. We adopt two common ranking evaluation methods: Accuracy@$k$ and Mean Reciprocal Rank (MRR), to evaluate the performance of predicted ranking lists for users. Accuracy@$k$ was widely adopted by many previous ranking based recommender systems~\cite{koren2008factorization}, and MRR was also widely chosen as a metric for ranking tasks~\cite{voorhees1999trec}.  

The basic idea of Accuracy@$k$ is to test whether a user's favorite item appears in the predicted top-$k$ items list, we regard the positive items with positive feedbacks as the user's favorite items. For each positive feedback $r_{ui}\in D_{test}$: (1) we randomly choose 1000 negative items and compute predicted rating scores for the ground-truth item $i$ as well as  the 1000 negative items; (2) we form a ranked list by ordering these items according to their predicted  ratings; (3) if the ground-truth item $i$ appears in the top-$k$ ranked list, we have a `hit'; otherwise, we  have a `miss'. 


Accuracy@$k$ has been widely used in evaluating recommendation accuracy by assessing the quality of the obtained top-$k$ items list. Accuracy@$k$ is formulated as:

\begin{equation}
\operatorname{Accuracy@\emph{k}}=\frac{\#hit@k}{\left| D_{test}\right|},
\end{equation}
where $\left|D_{test} \right|$ is the size of the test set, and $\#hit@k$ denotes the number of `hit' in the test set.

The Mean Reciprocal Rank (MRR)~\cite{voorhees1999trec} is to evaluate a ranking task that produces a list of responses to a query, ordered by probability of correctness. The reciprocal rank of a query response is the multiplicative inverse of the rank of the first correct answer. The mean reciprocal rank is the average of the reciprocal ranks of results for a query, MRR is defined as:
\begin{equation}
\text{MRR}=\frac{1}{\left| D_{test} \right|}\sum\nolimits_{r_{ui}\in D_{test}}{\frac{1}{rank(u,i)}},
\end{equation}
where $rank(u,i)$ is the position of item $i$ in the obtained top-$k$ items list for user $u$.

\subsubsection{Competing Baselines and Experimental Settings}\label{sec6-3}
The proposed DPH has three key components: the ranking-based objective, the DAE framework for incorporating content information, and the application of hash technique. Intuitively, the combination of all three components will improve the recommendation performance. Thus, we do ablation studies in Section \ref{sec5-3-5} to evaluate the effectiveness of each component in DPH. Besides, we choose two types of comparison methods as illustrated in Table \ref{tab:0}, content-aware recommender systems including CTR, CDL, and VBPR, and hashing-based recommender systems including DCF, DPR, and DDL, to evaluate the effectiveness of the proposed DPH regarding cold-start and data sparse settings. In the Table \ref{tab:0}, binary (\checkmark) denotes binary representations and binary ($\times$) represents continuous representations. Ranking ($\times$) means the loss function of the method is rating-based objective. Content ($\times$) denotes the method only use the rating information and content (\checkmark) presents the method use both the rating and content data. In our experiments, we use cross-validation method to tune the optimal hyper-parameters for DPH and the competing baselines by grid search.

\begin{table}
	\caption{Categories of baselines and DPH (the proposed model): `Hash' means hash-based recommendation methods and `UN-Hash' means continuous based methods.} 
	\centering
	\begin{tabular}{c|c|c|c c}
		\hline
		 Category & Methods & Binary & Ranking & Content\\
		 \hline
		\multirow{3}{*}{UN-Hash} &
	CTR & $\times$ &  $\times$ & \checkmark \\
		\cline{2-5}
	&	CDL & $\times$ &  $\times$ & \checkmark \\
			\cline{2-5}
	&	VBPR & $\times$ &  \checkmark & \checkmark \\
		\hline
				
			\multirow{3}{*}{Hash} &
		DCF & \checkmark  &  $\times$ &$\times$  \\
		\cline{2-5}
	&	DPR & \checkmark & \checkmark  &  $\times$ \\
		\cline{2-5}	
	&	DDL & \checkmark & $\times$  & \checkmark \\
		\cline{2-5}
	&	DPH & \checkmark &  \checkmark & \checkmark \\
		\hline
	\end{tabular}
	\label{tab:0}
\end{table}
For the proposed DPH, we search the optimal hyper parameters $\alpha$, $\beta$, and $\lambda$ from $[10^{-5}, \cdots, 10^{5}]$, and we then set the $\alpha=10^{-5}$, $\beta=10^{-3}$, $\lambda=20$. To be consistent with other baselines, we set the layer structure of DAE as $[8000, 200, 30]$. 

For the content-aware baseline VBPR, we set $\lambda_{\Theta}=10$; for CTR, we set $\lambda_{u}=0.1$, $\lambda_{v}=10$, $a=1$, $b=0.01$ and $K=30$ since it can achieve better performance; for CDL, we set $\lambda_{u}=1$, $\lambda_{v}=10$, $\lambda_{n}=1e^4$, $\lambda_{w}=1e^{-4}$, $a=1$, $b=0.01$, and set the layer structure of SDAE as $[8000, 200, 30]$ for aligning with the dimension of other methods. 

For the hashing baselines DCF, we search the optimal hyper parameters $\alpha$, $\beta$, and $\lambda$ from $[10^{-5}, \cdots, 10^{5}]$, and set $\alpha=10^{-3}$, $\beta=10^{-3}$; For DPR, we set $\alpha=10^{-4}$, $\beta=10^{-3}$; for DDL, we set $\lambda=5$, $\alpha=10^{-3}$, $\beta=10^{-3}$, and the layer structure of DBN as $[8000, 200, 30]$.


\subsection{Experimental Results and Analysis}\label{sec7}
we evaluate the effectiveness of the proposed DPH in three aspects: 
\begin{itemize}
	\item the superiority over competing baselines in cold-start settings.
	\item the competing performance in various sparse settings (sparsity levels of 10\%, 20\%, 30\%, and 40\%), 
	\item the high efficiency performance for online recommendation. 
\end{itemize}
\begin{figure}[t]
	\centering
	\includegraphics[width=1\linewidth]{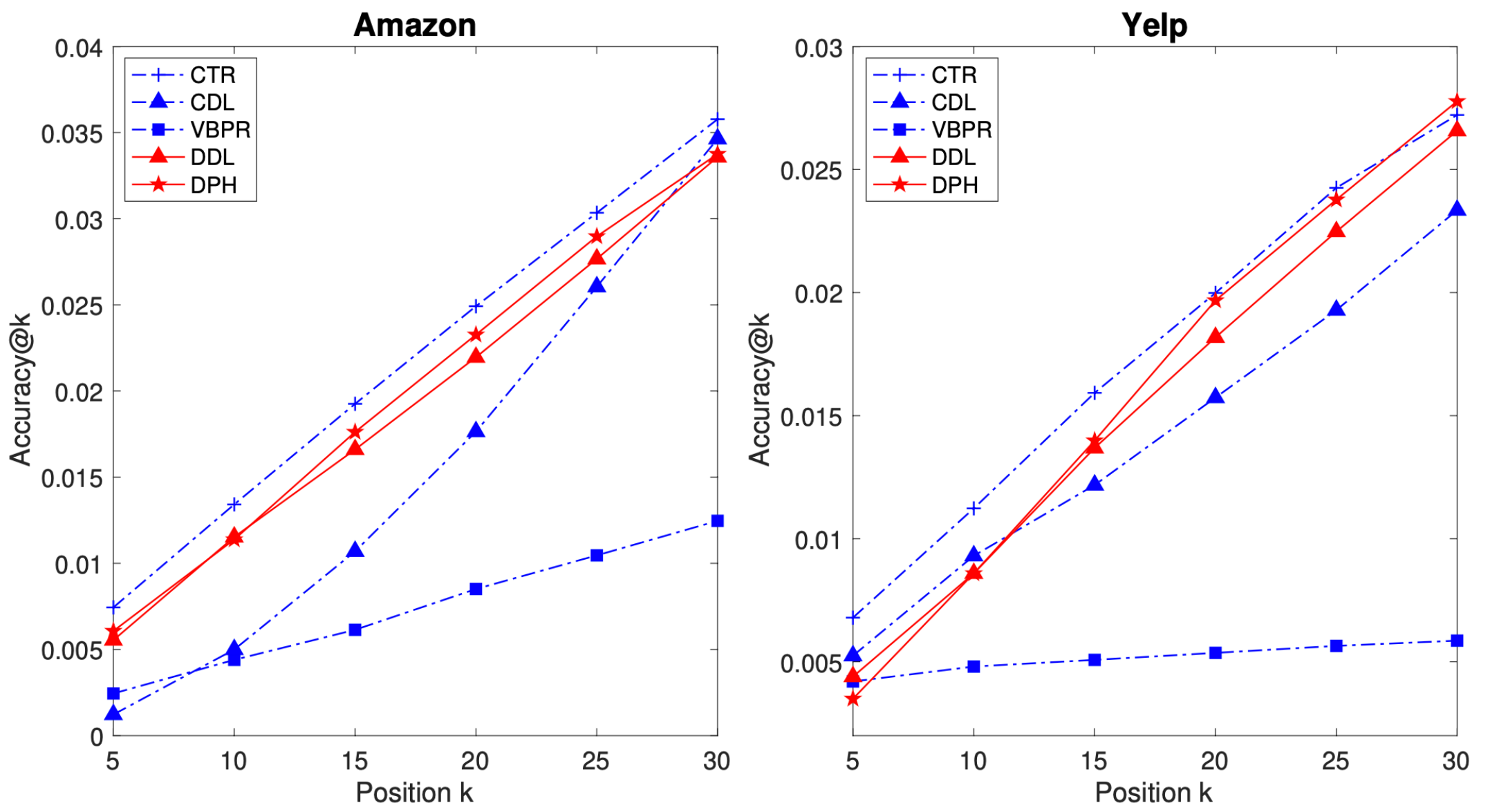}
	\caption{Recommendation accuracy comparison in cold-start setting with sparsity level of 10\%.}\label{fig3}
\end{figure}

\begin{figure}[t]
	\centering
\includegraphics[width=1\linewidth]{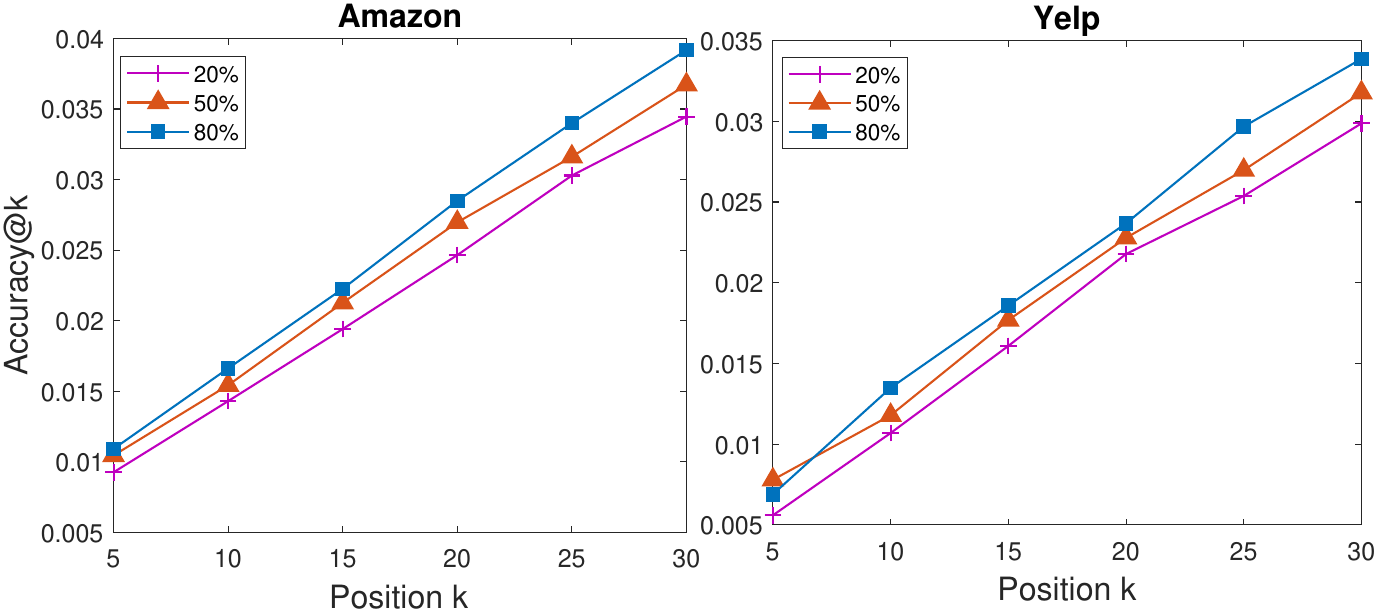}
	\caption{Recommendation accuracy of DPH varies with sparsity levels in cold-start setting.}\label{fig4}
\end{figure}

\begin{figure*}[t]
	\centering

  \includegraphics[width=1\linewidth]{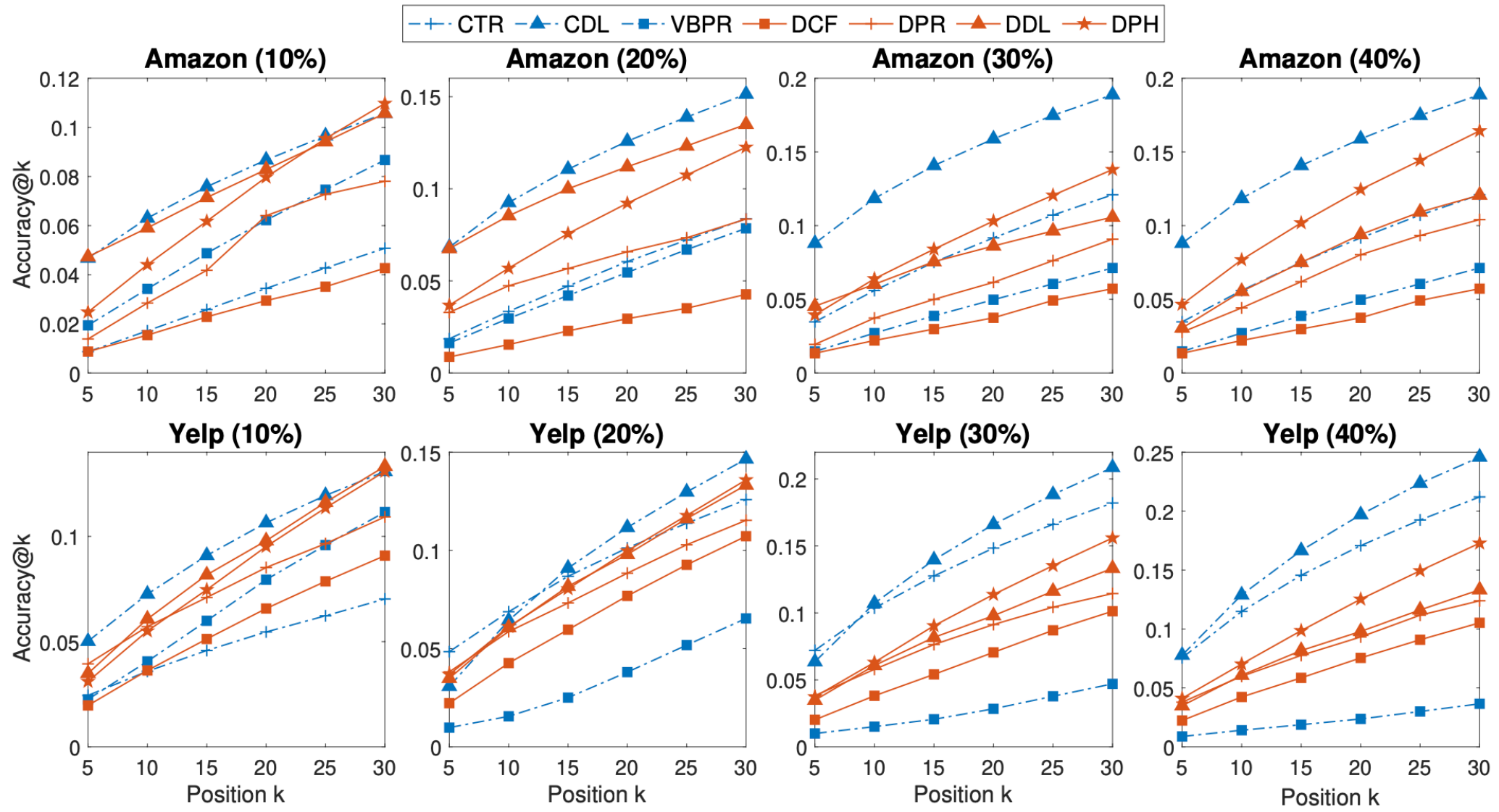}
\caption{Recommendation accuracy comparison on Amazon and Yelp datasets w.r.t different sparsity levels.}\label{fig5}
\end{figure*}

\begin{figure}[t]
	\centering

    \includegraphics[width=1\linewidth]{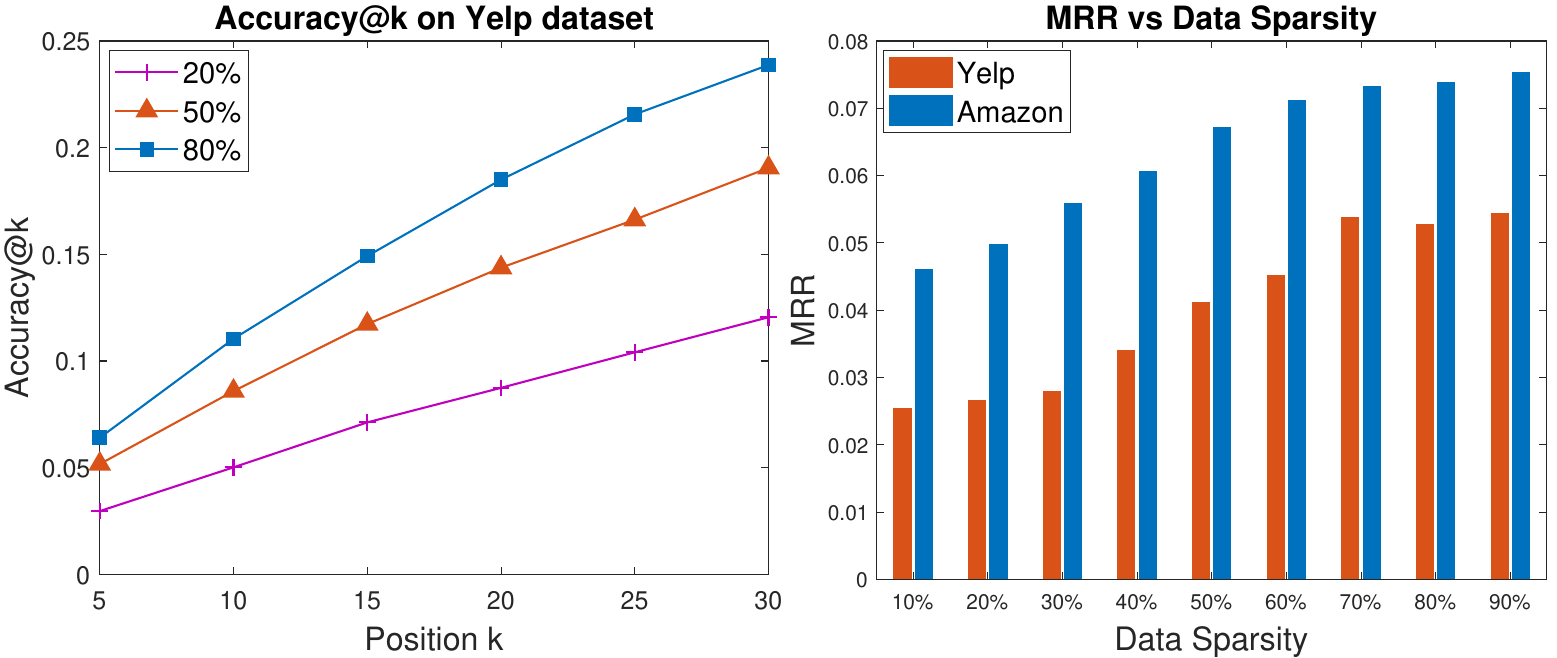}
    \caption{Recommendation accuracy of DPH varies with sparsity levels in sparse settings.}\label{fig6}
\end{figure}

\subsubsection{Recommendation Accuracy in Cold-start Setting}\label{sec7-1}
In this section, we compare the proposed DPH with three content-aware competing baselines and one hashing-based baseline, separately drawn as blue dashed lines and solid red lines in the following figures.

To evaluate the superior performance over other baselines in cold-start item setting, we first choose items with less than five ratings (positive feedbacks) as cold-start items, and test the performance on these cold-start items' rating data, denoted as $D_{test}^{c}$. To train our model, we randomly select 10\% rating data as training data, $D_{train}$, from ratings of all users and items except for cold-start ones. We then test the performance on the simulated cold-start items' ratings $D_{test}^{c}$, and other remaining ratings $D_{test}^{s}$, and separately report the results as performances in cold-start and sparse settings.

We adopt Accuracy@k and MRR to respectively evaluate recommendation accuracy in the clod-start setting with the sparsity of 10\%. The accuracy@k comparison of Amazon and Yelp datasets shown in Figure~\ref{fig3} indicates the superiority of DPH over the other hashing based baseline DDL on two datasets, as well as the competing performance compared with real-valued content-aware baselines on Amazon dataset, plus better performance than all baselines on Yelp dataset. The real-valued content-aware recommender systems can naturally achieve better accuracy than hashing based recommender systems, due to real-valued vectors intuitively carried much more information than hash codes. Because each dimension of a real-valued vector is stored with 32/64 bits, while each dimension of a hash code is saved by only one bit. That's the reason why hashing-based recommendation has inferior accuracy performance to the real-valued recommendation. But hashing-based frameworks have a significant advantage in online recommendation over real-valued methods, which will be evaluated in Section~\ref{sec7-3}. Thus it is acceptable and reasonable to have small accuracy gaps between real-valued content-aware recommender systems and the proposed hashing-based DPH.

The MRR comparison displayed in Table~\ref{tab:3} summarizes MRR results of the proposed DPH and four baselines, the best result is marked as \lq${\star}$\rq~ and the second best is marked as~\lq${o}$\rq~. We can conclude that the performance of DPH is very close to the best result, that is consistent with the result of Accuracy@$k$. We can also explore DDL and CDL can also perform well in cold-start setting, because we use a similar idea with the combination of deep representation learning and collaborative filtering. But we differ in designing objectives: CDL and DDL are rating based objectives, while DPH is pairwise ranking based objective, and the experiments in cold-start setting regarding the two above metrics show its superiority over other baselines.  

Figure~\ref{fig4} reveals the recommendation accuracy in clod-start settings varies with respect to sparsity levels. As discussed at the beginning of Section~\ref{sec7-1}, the recommendation performance on $D_{test}^{c}$ is dependent on the sparsity level. Because we can learn better users' representation as well as better DAE structure (including parameters), if we apply more ratings for training. The result shown in Figure~\ref{fig4} also evaluate the intuitive observation on Amazon and Yelp datasets, respectively.

\begin{table}
  \caption{MRR in cold-start item setting (with sparsity level of 10\%).}
  \centering
  \begin{tabular}{c|c|c|c|c|c}
  \hline
  \textbf{Methods} & \textbf{CTR} & \textbf{CDL} & \textbf{VBPR} & \textbf{DDL}&\textbf{DPH} \\
  \hline
  Amazon &0.0082$^{\star}$ & 0.0071 & 0.0042 & 0.0077 & 0.0079$^{o}$ \\
    \hline
  Yelp & 0.0059 &   0.0022 &  0.0067$^{o}$ &  0.0060 & 0.0074$^{\star}$  \\
  \hline
 \end{tabular}
 \label{tab:3}
\end{table}

\subsubsection{Recommendation Accuracy in Sparse Settings}
Figure~\ref{fig4} reveals the recommendation accuracy in cold-start settings varies concerning sparsity levels. As discussed at the beginning of Section~\ref{sec7-1}, the recommendation performance on $D_{test}^{c}$ is dependent on the sparsity level. We can learn better users' representation as well as better DAE structure (including parameters) if we apply more ratings for training. The result shown in Figure~\ref{fig4} also evaluate the intuitive observation on Amazon and Yelp datasets, respectively.

\begin{table}
  \caption{MRR w.r.t different sparsity levels (10\%, 20\% ).} 
  \centering
  \begin{tabular}{c|c c|c c}
  \hline
  \multirow{2}{*}{} &
 \multicolumn{2}{c|}{Amazon} &
 \multicolumn{2}{c}{Yelp} \\
 \cline{2-5}
   & 10\% & 20\% &  10\% & 20\% \\
  \hline
  \textbf{CTR} &0.0102	&0.0168	&0.0194&0.0355$^{o}$ \\
  \hline
  \textbf{CDL} & 0.0311	&0.0481$^{o}$ & 0.0360$^{\star}$ &0.0558$^{\star}$\\
  \hline
  \textbf{VBPR} &0.0176  &0.0158	&0.0206 &0.0126 \\
  \hline
    \textbf{DCF}&0.0102	&0.0103	&0.0195 &0.0221 \\
  \hline
   \textbf{DPR}&0.0200	&0.0317	&0.1581 &0.1161 \\
  \hline
  \textbf{DDL}&0.0490$^{\star}$	&0.0263	&0.0275 &0.0151 \\
  \hline
  \textbf{DPH}&0.0460$^{o}$	&0.050$^{\star}$ &0.0279$^{o}$ & 0.0296 \\
  \hline
 \end{tabular}
 \label{tab:4}
\end{table}
%
%
%
%

\subsubsection{Comparison with Hashing-based Frameworks}
From Figure~\ref{fig5}, we can see that DPH outperforms other hashing-based baselines: DCF, DPR, and DDL. By comparison of DDL, the Accuracy@k of DPH increases steadily with the increasing ratings (from 10\% to 40\%) for training, while the performance of DDL is not stable. From the observation for the performance of DDL, CDL, and together with DPH, we consider that the robust deep learning framework DAE applied in DPH works well on two datasets with different sparsity levels. As for DCF and DPR, they did not combine the content data into the collaborative filtering frameworks, which leads to poor performance in sparse settings. 
\subsubsection{Accuracy in Various Sparsity Settings}
By the analysis of Figure~\ref{fig4} and Figure~\ref{fig6}, we can conclude the recommendation performances in cold-start, and sparse settings are influenced by the sparsity of training data. Training with more ratings learns better users' representation and a better deep structure, which is helpful to conduct a cold-start item recommendation, and beneficial to the sparse recommendation.

\subsubsection{Ablation Study} \label{sec5-3-5}
Compared with DDL, the proposed DPH adopts DAE other than DBN to integrate content embedding, and it is formulated with the ranking-based objective instead of the rating-based objective. Thus, we do some ablation studies to show the effectiveness of the above two components. Figure 7 demonstrates the performance comparison in terms of the metric Accuracy@k with DPH-R, DPH-B, DPH-B-R on Amazon and Yelp (with 30\% sparsity). The only difference between DPH-R and DPH is that DPH-R is modeled as the rating-based objective; while DPH is formulated as the ranking-based objective. The only difference between DPH-B and DPH is that DPH-B integrates content information by DBN but DPH combines content data with DAE. There are two differences between DPH-B-R and DPH: DPH-B-R is formulated with the rating-based objective and it exploits content data with the DBN framework. 

Figure 7 tells us that the performance of DPH is almost the best on both two datasets, and DPH-B-R has competitive performance with DPH. Actually, the DPH-B-R is DDL, and it performs better than other two baselines DPH-B and DPH-R, which tells us that the combination of the rating-based objective and the DBN framework is better than other two combinations (DPH-B and DPH-R). Because the rating-based objective model reconstructs ratings itself which is consistent with the goal of DBN framework that reconstructs each layer's entire input as accurately as impossible. DPH performs the best with the combination of the ranking-based objective and the DAE framework, because the ranking-based objective models the groundtruth of ranking which is in line with the ultimate goal of top-k recommendation. Besides, the DAE framework not only preserves the input information, but it also eliminates the effect of a corruption process stochastically applied to the input of the autoencoder, and it forces the hidden layer to discover more robust embeddings.

\begin{figure}[t]
	\centering	
	\includegraphics[width=1\linewidth]{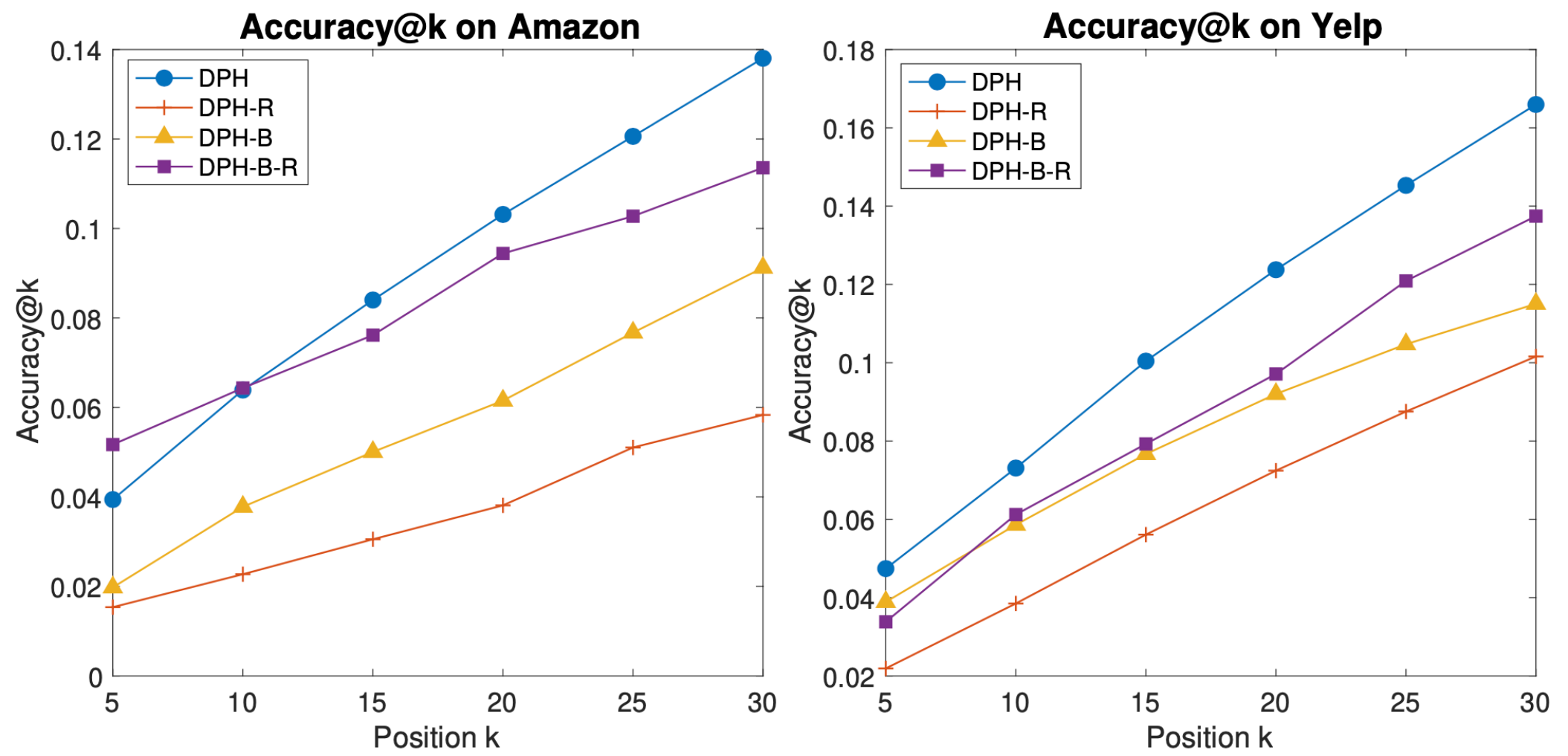}
	\caption{Ablation studies with rating-based (`R') or ranking-based objectives, DBN (`B') or DAE for integrating content information (with 30\% sparsity).}\label{fig8}
\end{figure}

\subsubsection{Efficiency for Online Recommendation}\label{sec7-3}
Compared with real-valued recommender systems, hashing-based frameworks has significant advantage of providing efficient online recommendation. We separately evaluate the efficiency regarding time and storage on synthetic data.

\textbf{Time Complexity:} In real-valued recommender systems, users and items are denoted by $d-$dimension real-valued vectors; while in hashing-based recommender systems, users and items are represented by $d-$dimension hash codes. For real-valued methods, the time cost of finding top$-k$ items from $m$ items is $\mathcal{O}(md + m\log k )$ for each user $u$; while for hashing methods, searching nearest neighbors in Hamming space is extremely fast. There are two methods to search the top-$k$ items: one is Hamming ranking, it can be conducted by ranking Hamming distances with the query hash code (user), it has the complexity of $\mathcal{O}(m)$, which is linear with the item data size. The other one is hashing lookup. Similar hash codes are searched in a hamming ball centered at the query hash code. The time complexity is independent of the item data size. Therefore, Hashing based recommendation has evident superiority over the real-valued recommendation.

\begin{figure}[t]
	\centering	
    \includegraphics[width=1\linewidth]{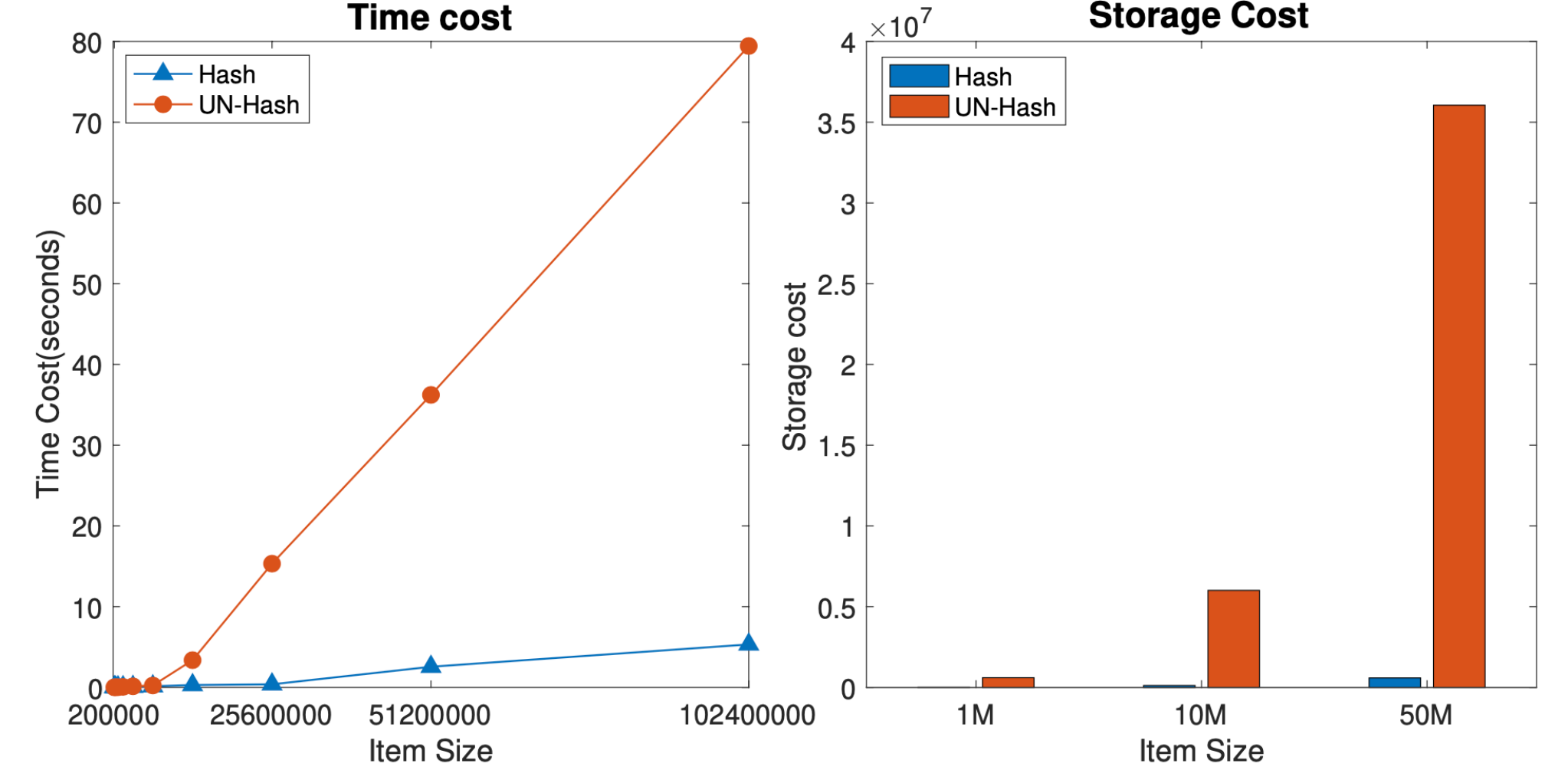}
	\caption{Efficiency comparison between `UN-Hash' (continuous) methods and `Hash' methods for online recommendation on synthetic data. Left: time cost comparison. Right: storage cost comparison.}\label{fig7}
\end{figure}

In this paper, we investigate the time cost comparison on synthetic datasets. We first use standard Gaussian distribution to generate users' and items' real-valued features randomly. Items hash codes are obtained from real-valued vectors by the sign function. We set different sizes of items sets in the experiment: 200,000, 400,000, ... , 102,400,000, to test the time cost variation of online recommendation. The variation is shown in the left of Figure~\ref{fig7}. In addition, we evaluate the time efficiency on Amazon and Yelp datasets as well in the Table \ref{tab:5}, where `UN-Hash' denotes continuous based recommendations and `Hash' represents hashing based methods. Experimental results on synthetic and real world datasets tells us that the time cost of real-valued vectors grows fast with the item number, in comparison, the time cost of hash codes increases much slower than continuous features. The experimental results show that hashing based recommendation has an evident advantage over the real-valued recommendation for online recommendation.

\begin{table}
	\caption{Time cost comparison on Amazon and Yelp ($\times 10^{-4}$seconds).}
	\centering
	\begin{tabular}{c|c|c}
		\hline
		\textbf{Methods} & 	\textbf{UN-Hash}&	\textbf{Hash} \\
		\hline
	 \textbf{Amazon} &4.0491 & 1.0971\\
		\hline
	 \textbf{Yelp} & 12.5975  &  1.9878  \\
		\hline
	\end{tabular}
	\label{tab:5}
\end{table}

\textbf{Storage Complexity:} In the Real space, at least 64 bits are needed to store one dimension of a vector. As the dimension becomes large, it will cost much more space. While in Hamming space, only one bit is needed to store one dimension of a hash code. Thus the storage cost is reduced significantly.

We test the stdorage costs of hash codes and real-valued vectors on three different sizes of item sets: 1 million, 10 million, and 50 million. In the right of Figure~\ref{fig7}, `UN-Hash' represents continuous based methods and `Hash' denotes hashing based models. It says that hash codes obtained from hashing models cost much less memory to store the same number of items.
\section{Conclusion}\label{sec8}
In this paper, we propose a robust hashing-based recommendation framework deep pairwise hashing (DPH) to cope with the cold-start and sparse issues by integrating content and rating information. Specifically, to align with the ultimate goal of recommender system, we formulate the proposed framework DPH with a ranking-based objective to improve the performance. Besides, we choose Denoising Auto-encoder to extract robust items' representations from the content data. Then, we add uncorrelated and independent constraints on hash codes to learn short and informative hash codes. Furtherly, we develop an alternating optimization algorithm together with a discrete coordinate descent method to train the proposed model. Finally, we evaluate the effectiveness of the proposed DPH by metrics Accuracy@k and MRR on Amazon and Yelp datasets. Experimental results show its consistent superiority over the competing hashing-based baselines in cold-start item and sparse settings, and also demonstrate its competing performance with the continuous based content-aware recommender systems.
\section{Acknowledgment}
This work was supported by National Natural Science Foundation of China (Grant No. 61572109, 61801060). Prof.ni Ivor W. Tsang was supported by ARC FT130100746, DP180100106 and LP150100671. Dr. Hongzhi Yin was supported by ARC DE160100308, DP170103954, and New Staff Research Grant of the University of Queensland (Grant No.613134).
\bibliographystyle{IEEEtran}
\bibliography{tkde}

\begin{thebibliography}{10}
\providecommand{\url}[1]{#1}
\csname url@samestyle\endcsname
\providecommand{\newblock}{\relax}
\providecommand{\bibinfo}[2]{#2}
\providecommand{\BIBentrySTDinterwordspacing}{\spaceskip=0pt\relax}
\providecommand{\BIBentryALTinterwordstretchfactor}{4}
\providecommand{\BIBentryALTinterwordspacing}{\spaceskip=\fontdimen2\font plus
\BIBentryALTinterwordstretchfactor\fontdimen3\font minus
  \fontdimen4\font\relax}
\providecommand{\BIBforeignlanguage}[2]{{%
\expandafter\ifx\csname l@#1\endcsname\relax
\typeout{** WARNING: IEEEtran.bst: No hyphenation pattern has been}%
\typeout{** loaded for the language `#1'. Using the pattern for}%
\typeout{** the default language instead.}%
\else
\language=\csname l@#1\endcsname
\fi
#2}}
\providecommand{\BIBdecl}{\relax}
\BIBdecl

\bibitem{schein2002methods}
A.~I. Schein, A.~Popescul, L.~H. Ungar, and D.~M. Pennock, ``Methods and
  metrics for cold-start recommendations,'' in \emph{Proceedings of the 25th
  annual international ACM SIGIR conference on Research and development in
  information retrieval}.\hskip 1em plus 0.5em minus 0.4em\relax ACM, 2002, pp.
  253--260.

\bibitem{wang2011collaborative}
C.~Wang and D.~M. Blei, ``Collaborative topic modeling for recommending
  scientific articles,'' in \emph{Proceedings of the 17th ACM SIGKDD
  international conference on Knowledge discovery and data mining}.\hskip 1em
  plus 0.5em minus 0.4em\relax ACM, 2011, pp. 448--456.

\bibitem{wang2015collaborative}
H.~Wang, N.~Wang, and D.-Y. Yeung, ``Collaborative deep learning for
  recommender systems,'' in \emph{Proceedings of the 21th ACM SIGKDD
  International Conference on Knowledge Discovery and Data Mining}.\hskip 1em
  plus 0.5em minus 0.4em\relax ACM, 2015, pp. 1235--1244.

\bibitem{zhang2016collaborative}
F.~Zhang, N.~J. Yuan, D.~Lian, X.~Xie, and W.-Y. Ma, ``Collaborative knowledge
  base embedding for recommender systems,'' in \emph{Proceedings of the 22nd
  ACM SIGKDD International Conference on Knowledge Discovery and Data
  Mining}.\hskip 1em plus 0.5em minus 0.4em\relax ACM, 2016, pp. 353--362.

\bibitem{wu2016collaborative}
Y.~Wu, C.~DuBois, A.~X. Zheng, and M.~Ester, ``Collaborative denoising
  auto-encoders for top-n recommender systems,'' in \emph{Proceedings of the
  Ninth ACM International Conference on Web Search and Data Mining}, 2016, pp.
  153--162.

\bibitem{yin2014lcars}
H.~Yin, B.~Cui, Y.~Sun, Z.~Hu, and L.~Chen, ``Lcars: A spatial item recommender
  system,'' \emph{ACM Transactions on Information Systems (TOIS)}, vol.~32,
  no.~3, p.~11, 2014.

\bibitem{yin2016adapting}
H.~Yin, X.~Zhou, B.~Cui, H.~Wang, K.~Zheng, and Q.~V.~H. Nguyen, ``Adapting to
  user interest drift for poi recommendation,'' \emph{IEEE Transactions on
  Knowledge and Data Engineering}, vol.~28, no.~10, pp. 2566--2581, 2016.

\bibitem{zhou2012learning}
K.~Zhou and H.~Zha, ``Learning binary codes for collaborative filtering,'' in
  \emph{Proc. of ACM SIGKDD}.\hskip 1em plus 0.5em minus 0.4em\relax ACM, 2012,
  pp. 498--506.

\bibitem{zhang2016discrete}
H.~Zhang, F.~Shen, W.~Liu, X.~He, H.~Luan, and T.-S. Chua, ``Discrete
  collaborative filtering,'' in \emph{Proceedings of the 39th International ACM
  SIGIR conference on Research and Development in Information Retrieval}.\hskip
  1em plus 0.5em minus 0.4em\relax ACM, 2016, pp. 325--334.

\bibitem{zhang2017discrete}
Y.~Zhang, D.~Lian, and G.~Yang, ``Discrete personalized ranking for fast
  collaborative filtering from implicit feedback,'' in \emph{Thirty-First AAAI
  Conference on Artificial Intelligence}, 2017.

\bibitem{wang2012semi}
J.~Wang, S.~Kumar, and S.-F. Chang, ``Semi-supervised hashing for large-scale
  search,'' \emph{IEEE TPAMI}, vol.~34, no.~12, pp. 2393--2406, 2012.

\bibitem{rendle2009bpr}
S.~Rendle, C.~Freudenthaler, Z.~Gantner, and L.~Schmidt-Thieme, ``Bpr: Bayesian
  personalized ranking from implicit feedback,'' in \emph{Proceedings of
  UAI'09}.\hskip 1em plus 0.5em minus 0.4em\relax AUAI Press, 2009, pp.
  452--461.

\bibitem{weimer2008cofi}
M.~Weimer, A.~Karatzoglou, Q.~V. Le, and A.~J. Smola, ``Cofi rank-maximum
  margin matrix factorization for collaborative ranking,'' in \emph{Advances in
  neural information processing systems}, 2008, pp. 1593--1600.

\bibitem{shi2010list}
Y.~Shi, M.~Larson, and A.~Hanjalic, ``List-wise learning to rank with matrix
  factorization for collaborative filtering,'' in \emph{Proceedings of
  RecSys'10}.\hskip 1em plus 0.5em minus 0.4em\relax ACM, 2010, pp. 269--272.

\bibitem{zhang2018discrete}
Y.~Zhang, H.~Yin, Z.~Huang, X.~Du, G.~Yang, and D.~Lian, ``Discrete deep
  learning for fast content-aware recommendation,'' in \emph{Proceedings of the
  Eleventh ACM International Conference on Web Search and Data Mining}.\hskip
  1em plus 0.5em minus 0.4em\relax ACM, 2018, pp. 717--726.

\bibitem{salakhutdinov2007probabilistic}
R.~Salakhutdinov and A.~Mnih, ``Probabilistic matrix factorization.'' in
  \emph{Nips}, vol.~1, no.~1, 2007, pp. 2--1.

\bibitem{zhang2019deep}
S.~Zhang, L.~Yao, A.~Sun, and Y.~Tay, ``Deep learning based recommender system:
  A survey and new perspectives,'' \emph{ACM Computing Surveys (CSUR)},
  vol.~52, no.~1, pp. 1--38, 2019.

\bibitem{vincent2010stacked}
P.~Vincent, H.~Larochelle, I.~Lajoie, Y.~Bengio, and P.-A. Manzagol, ``Stacked
  denoising autoencoders: Learning useful representations in a deep network
  with a local denoising criterion,'' \emph{Journal of Machine Learning
  Research}, vol.~11, no. Dec, pp. 3371--3408, 2010.

\bibitem{he2016vbpr}
R.~He and J.~McAuley, ``Vbpr: visual bayesian personalized ranking from
  implicit feedback,'' in \emph{Thirtieth AAAI Conference on Artificial
  Intelligence}, 2016.

\bibitem{das2007google}
A.~S. Das, M.~Datar, A.~Garg, and S.~Rajaram, ``Google news personalization:
  scalable online collaborative filtering,'' in \emph{Proc. of WWW}.\hskip 1em
  plus 0.5em minus 0.4em\relax ACM, 2007, pp. 271--280.

\bibitem{datar2004locality}
M.~Datar, N.~Immorlica, P.~Indyk, and V.~S. Mirrokni, ``Locality-sensitive
  hashing scheme based on p-stable distributions,'' in \emph{Proceedings of the
  twentieth annual symposium on Computational geometry}.\hskip 1em plus 0.5em
  minus 0.4em\relax ACM, 2004, pp. 253--262.

\bibitem{karatzoglou2010collaborative}
A.~Karatzoglou, M.~Weimer, and A.~J. Smola, ``Collaborative filtering on a
  budget,'' in \emph{International Conference on Artificial Intelligence and
  Statistics}, 2010, pp. 389--396.

\bibitem{gong2013iterative}
Y.~Gong, S.~Lazebnik, A.~Gordo, and F.~Perronnin, ``Iterative quantization: A
  procrustean approach to learning binary codes for large-scale image
  retrieval,'' \emph{IEEE Transactions on Pattern Analysis and Machine
  Intelligence}, vol.~35, no.~12, pp. 2916--2929, 2013.

\bibitem{zhang2014preference}
Z.~Zhang, Q.~Wang, L.~Ruan, and L.~Si, ``Preference preserving hashing for
  efficient recommendation,'' in \emph{Proc. of SIGIR}.\hskip 1em plus 0.5em
  minus 0.4em\relax ACM, 2014, pp. 183--192.

\bibitem{zhang2017dot}
Y.~Zhang, G.~Yang, L.~Hu, H.~Wen, and J.~Wu, ``Dot-product based preference
  preserved hashing for fast collaborative filtering,'' in \emph{Communications
  (ICC), 2017 IEEE International Conference on}.\hskip 1em plus 0.5em minus
  0.4em\relax IEEE, 2017, pp. 1--6.

\bibitem{vincent2008extracting}
P.~Vincent, H.~Larochelle, Y.~Bengio, and P.-A. Manzagol, ``Extracting and
  composing robust features with denoising autoencoders,'' in \emph{Proceedings
  of the 25th international conference on Machine learning}.\hskip 1em plus
  0.5em minus 0.4em\relax ACM, 2008, pp. 1096--1103.

\bibitem{bradley1997use}
A.~P. Bradley, ``The use of the area under the roc curve in the evaluation of
  machine learning algorithms,'' \emph{Pattern recognition}, vol.~30, no.~7,
  pp. 1145--1159, 1997.

\bibitem{koren2011advances}
Y.~Koren and R.~Bell, ``Advances in collaborative filtering,'' in
  \emph{Recommender systems handbook}.\hskip 1em plus 0.5em minus 0.4em\relax
  Springer, 2011, pp. 145--186.

\bibitem{gao2013one}
W.~Gao, R.~Jin, S.~Zhu, and Z.-H. Zhou, ``One-pass auc optimization,'' in
  \emph{International Conference on Machine Learning}, 2013, pp. 906--914.

\bibitem{haastad2001some}
J.~H{\aa}stad, ``Some optimal inapproximability results,'' \emph{Journal of the
  ACM}, vol.~48, no.~4, pp. 798--859, 2001.

\bibitem{Shen_2015_CVPR}
F.~Shen, C.~Shen, W.~Liu, and H.~Tao~Shen, ``Supervised discrete hashing,'' in
  \emph{CVPR}, June 2015, pp. 37--45.

\bibitem{porter1980algorithm}
M.~F. Porter, ``An algorithm for suffix stripping,'' \emph{Program}, vol.~14,
  no.~3, pp. 130--137, 1980.

\bibitem{ramos2003using}
J.~Ramos \emph{et~al.}, ``Using tf-idf to determine word relevance in document
  queries,'' in \emph{Proceedings of the first instructional conference on
  machine learning}, vol. 242, 2003, pp. 133--142.

\bibitem{ma2008sorec}
H.~Ma, H.~Yang, M.~R. Lyu, and I.~King, ``Sorec: social recommendation using
  probabilistic matrix factorization,'' in \emph{Proceedings of the 17th ACM
  conference on Information and knowledge management}.\hskip 1em plus 0.5em
  minus 0.4em\relax ACM, 2008, pp. 931--940.

\bibitem{koren2008factorization}
Y.~Koren, ``Factorization meets the neighborhood: a multifaceted collaborative
  filtering model,'' in \emph{Proceedings of the 14th ACM SIGKDD international
  conference on Knowledge discovery and data mining}.\hskip 1em plus 0.5em
  minus 0.4em\relax ACM, 2008, pp. 426--434.

\bibitem{voorhees1999trec}
E.~M. Voorhees \emph{et~al.}, ``The trec-8 question answering track report,''
  in \emph{Trec}, vol.~99, 1999, pp. 77--82.

\end{thebibliography}

\begin{IEEEbiography}[{\includegraphics[width=1in,height=1.25in,clip,keepaspectratio]{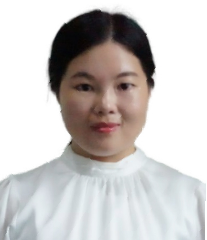}}]{Yan Zhang}
	received her BSc degree in mathematics and applied mathematics from Sichuan normal university, China. She is currently working toward the PhD degree in School of Computer Science and Engineering, University of Electronic Science and Technology of China, China. Her research interests include recommender system, machine learning, and logic synthesis.
\end{IEEEbiography}
\vspace{-15 mm}
\begin{IEEEbiography}[{\includegraphics[width=1in,height=1.25in,clip,keepaspectratio]{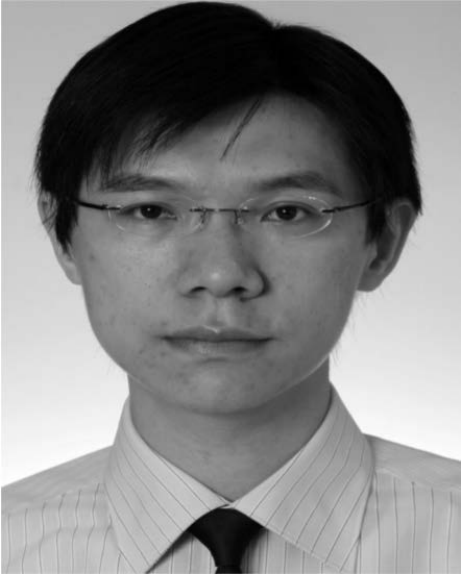}}]{Ivor W. Tsang} is an ARC Future Fellow and Professor of Artificial Intelligence, at University of Technology Sydney (UTS). He is also the Research Director of the UTS Flagship Research Centre for Artificial Intelligence (CAI). According to Google Scholar, he has more than 12,000 citations and his H-index is 53. In 2009, Prof Tsang was conferred the 2008 Natural Science Award (Class II) by Ministry of Education, China, which recognized his contributions to kernel methods. In 2013, Prof Tsang received his prestigious Australian Research Council Future Fellowship for his research regarding Machine Learning on Big Data. In addition, he had received the prestigious IEEE Transactions on Neural Networks Outstanding 2004 Paper Award in 2007, the 2014 IEEE Transactions on Multimedia Prize Paper Award, and the Best Student Paper Award at CVPR 2010. He serves as an Associate Editor for the IEEE Transactions on Big Data, the IEEE Transactions on Emerging Topics in Computational Intelligence and Neurocomputing, and area chair for NeurIPS.
\end{IEEEbiography}
\vspace{-15 mm}
\begin{IEEEbiography}[{\includegraphics[width=1in,height=1.25in,clip,keepaspectratio]{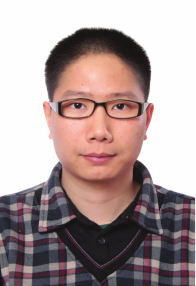}}]{Hongzhi Yin} received the PhD degree in computer science from Peking University, in 2014. He is a senior lecturer with the University of Queensland. He won the Australia Research Council Discovery Early-Career Researcher Award in 2015. His research interests include recommendation system, user profiling, topic models, deep learning, social media mining, and location-based services.
\end{IEEEbiography}
\vspace{-15 mm}
\begin{IEEEbiography}[{\includegraphics[width=1in,height=1.25in,clip,keepaspectratio]{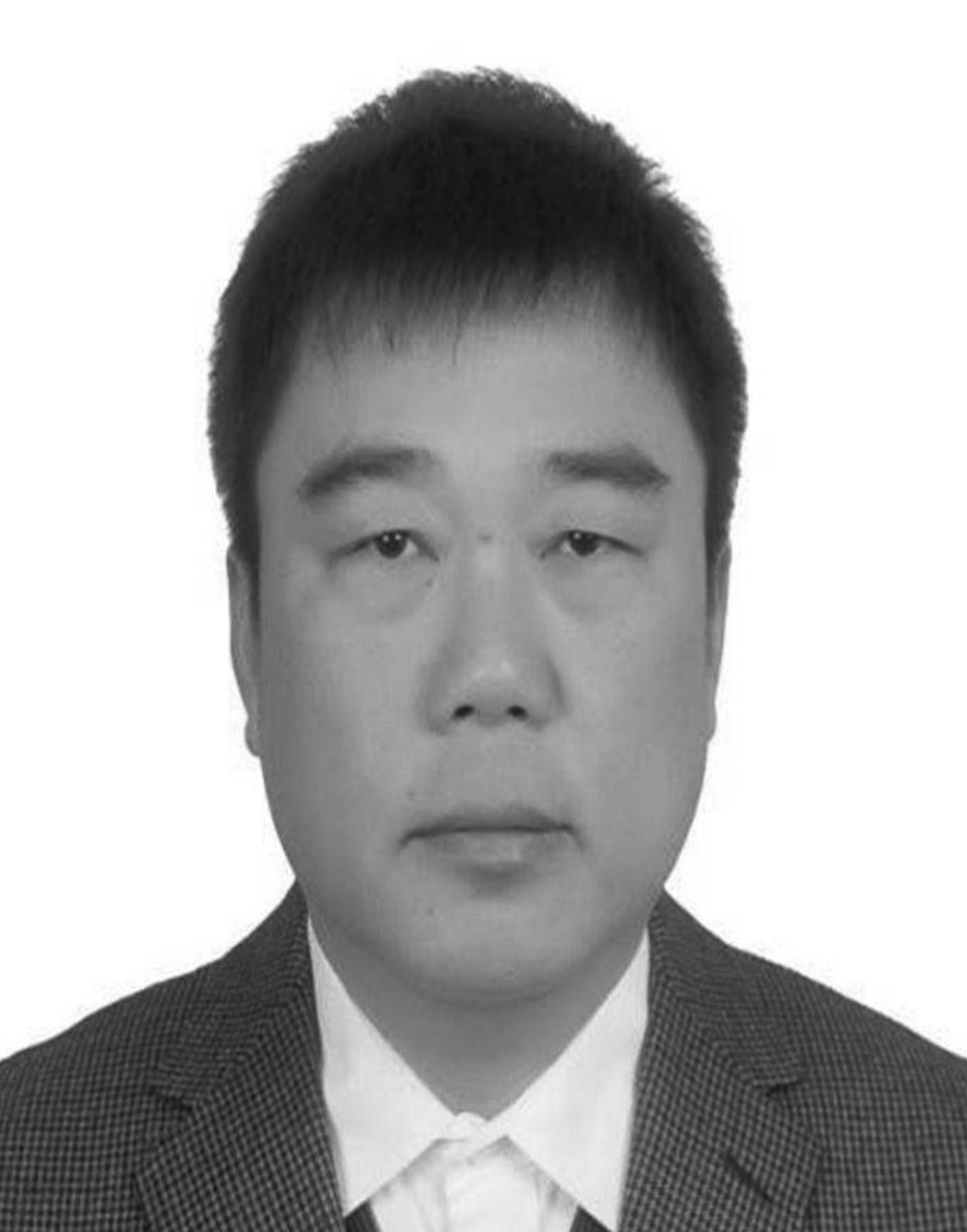}}]{Guowu Yang} received his BS degree from University of Science and Technology of China in 1989, received his MSc degree from Wuhan University of Technology in 1994, and received his PhD degree in electrical and computer engineering from Portland State University in 2005. He has worked at Wuhan University of Technology from 1989 to 2001, at Portland State University from 2005 to 2006. He is a full professor at University of Electronic Science and Technology of China now. His research interests include verification, logic synthesis, quantum computing and machine learning. He has published over 100 journal and conference papers.
\end{IEEEbiography}
\vspace{-15 mm}
\begin{IEEEbiography}[{\includegraphics[width=1in,height=1.25in,clip,keepaspectratio]{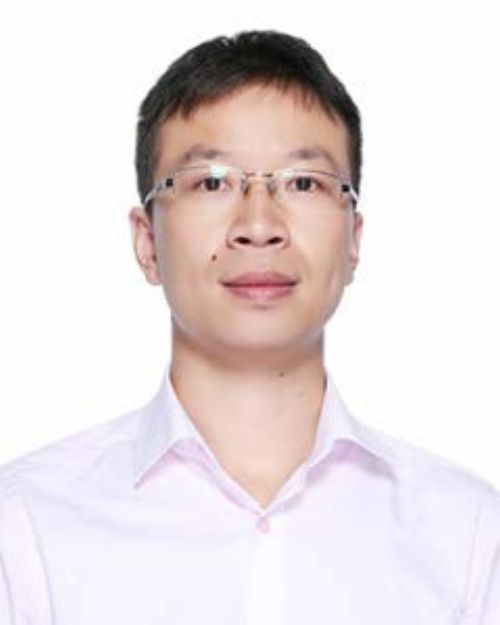}}]{Defu Lian} is a research professor in the School of Computer Science and Technology, University of Science and Technology of China (USTC), Hefei. He received the B.E. and Ph.D. degrees in computer science from University of Science and Technology of China (USTC) in 2009 and 2014, respectively. His research interest includes spatial data mining, recommender systems, and learning to hash.
\end{IEEEbiography}
\vspace{-15 mm}
\begin{IEEEbiography}[{\includegraphics[width=1in,height=1.25in,clip,keepaspectratio]{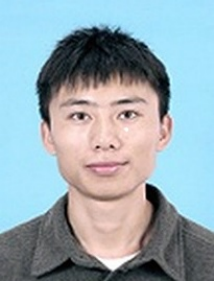}}]{Jingjing Li} received his MSc and PhD degree in Computer Science from University of Electronic Science and Technology of China in 2013 and 2017, respectively. Now he is a national Postdoctoral Program for Innovative Talents research fellow with the School of Computer Science and Engineering, University of Electronic Science and Technology of China. He has great interest in machine learning, especially transfer learning, subspace learning and recommender systems.
\end{IEEEbiography}
\end{document}